\documentclass{IEEEtran}
\usepackage{cite}
\usepackage{amsmath,amssymb,amsfonts}
\usepackage{algorithmic}
\usepackage{graphicx}
\usepackage{epstopdf}
\usepackage{textcomp}
\usepackage{flushend}
\def\BibTeX{{\rm B\kern-.05em{\sc i\kern-.025em b}\kern-.08em
    T\kern-.1667em\lower.7ex\hbox{E}\kern-.125emX}}
\AtBeginDocument{\definecolor{ojcolor}{cmyk}{0.93,0.59,0.15,0.02}}

\usepackage{booktabs}
\usepackage[usestackEOL]{stackengine}

\usepackage{mathtools}

\usepackage{multirow}
\usepackage{textcomp}
\usepackage{array}
\usepackage{tikz}
\usepackage{stfloats}
\usepackage{url}
\newcommand{\mathcenter}{\@fleqnfalse}

\begin{document}

\title{What Physical Layer Security Can Do for 6G Security}
\author{
	\IEEEauthorblockN{ Miroslav Mitev\IEEEauthorrefmark{1}, Arsenia Chorti\IEEEauthorrefmark{1},\IEEEauthorrefmark{2}, H. Vincent Poor\IEEEauthorrefmark{3}, Gerhard Fettweis\IEEEauthorrefmark{1},\IEEEauthorrefmark{4}}
	
 \IEEEauthorblockA{\IEEEauthorrefmark{1} Barkhausen Institut, Dresden, Germany; }
 
 \IEEEauthorblockA{\IEEEauthorrefmark{2} ETIS, UMR 8051 CY Cergy Paris Université, ENSEA, CNRS, Cergy, France; }
 
 \IEEEauthorblockA{\IEEEauthorrefmark{3} School of Engineering and Applied Science, Princeton University, Princeton, NJ, USA; }
 
 \IEEEauthorblockA{\IEEEauthorrefmark{4} Vodafone Chair for Mobile Communications Systems, Technische Universität Dresden, Dresden, Germany; \vspace{0.2cm} }

 \IEEEauthorblockA{Corresponding author: miroslav.mitev@barkhauseninstitut.org }
 }
% \author{Miroslav Mitev\authorrefmark{1}, Arsenia Chorti\authorrefmark{1}\authorrefmark{2}, Senior Member,~IEEE, H. Vincent Poor\authorrefmark{3}, 
% % , AND THIRD C. AUTHOR,~JR.\authorrefmark{1,2},
% Life Fellow, IEEE, Gerhard Fettweis\authorrefmark{1}\authorrefmark{4}, Fellow, IEEE}
% \affil{Barkhausen Institut, 01187 Dresden, Germany;}
% \affil{ETIS, UMR 8051 CY Cergy Paris Université, ENSEA, CNRS, 95000, Cergy, France;}
% \affil{School of Engineering and Applied Science, Princeton University, Princeton, NJ 08544, USA;}
% \affil{Vodafone Chair for Mobile Communications Systems, Technische Universität Dresden, 01062 Dresden , Germany;}
% \corresp{CORRESPONDING AUTHOR: Miroslav Mitev (e-mail: miroslav.mitev@barkhauseninstitut.org).}
% \authornote{The work is financed on the basis of the budget passed by the Saxon State Parliament. }
% \markboth{What Physical Layer Security Can Do for 6G Security}{M. Mitev \textit{et al.}}


%\markboth
%{Author \headeretal: Preparation of Papers for IEEE TRANSACTIONS and JOURNALS}
%{Author \headeretal: Preparation of Papers for IEEE TRANSACTIONS and JOURNALS}

% \corresp{Corresponding author: miroslav.mitev@barkhauseninstitut.org}


% \begin{keywords}
% 5G, 6G, physical layer security, wiretap coding, secret key generation, physical unclonable functions.
% \end{keywords}

% \titlepgskip=-15pt
\maketitle
\begin{abstract}
While existing security protocols were designed with a focus on the core network, the enhancement of the security of the B5G access network becomes of critical importance. Despite the strengthening of 5G security protocols with respect to LTE, there are still open issues that have not been fully addressed. This work is articulated around the premise that rethinking the security design bottom up, starting at the physical layer, is not only viable in 6G but importantly, arises as an efficient way to overcome security hurdles in novel use cases, notably massive machine type communications (mMTC), ultra reliable low latency communications (URLLC) and autonomous cyberphysical systems. Unlike existing review papers that treat physical layer security orthogonally to cryptography, we will try to provide a few insights of underlying connections. Discussing many practical issues, we will present a comprehensive review of the state-of the-art in i) secret key generation from shared randomness, ii)  the wiretap channels and fundamental limits, iii) authentication of devices using physical unclonable functions (PUFs), localization and multi-factor authentication, and, iv) jamming attacks at the physical layer. We finally conclude with the proposers’ aspirations for the 6G security landscape, in the hyper-connectivity and semantic communications era.
\end{abstract}
%\IEEEpeerreviewmaketitle
%%%%%%%%%%%%%%%%%%%%%%%%%%%%%%%%%%%%%%%%%%%%%%%%%%%%%%%%%%%%%%%%%%%%%%%%%%%%%%%%%%%%%%%%
%%%%%%%%%%%%%%%%%%%%%%%%%%%%%%%%%%%%%%%%%%%%%%%%%%%%%%%%%%%%%%%%%%%%%%%%%%%%%%%%%%%%%%%%
\section{Introduction} \label{sec:Introduction}
%%%%%%%%%%%%%%%%%%%%%%%%%%%%%%%%%%%%%%%%%%%%%%%%%%%%%%%%%%%%%%%%%%%%%%%%%%%%%%%%%%%%%%%%
%%%%%%%%%%%%%%%%%%%%%%%%%%%%%%%%%%%%%%%%%%%%%%%%%%%%%%%%%%%%%%%%%%%%%%%%%%%%%%%%%%%%%%%%
%Introduction structure:
%\begin{itemize}
%    \item Motivation
%    \item The problem
%    \item Other approaches to content-popularity detection, mainly model-based. They do not consider CPD, neither variance.
 %   \item Why CPD and variance change-point detection is important.
%    \item Our approach, what is novel. We do CPD, we consider variance, our algorithm is online, considers the most representative modeling approaches, applies more realistic modeling. 
%\end{itemize}
The rollout of fifth-generation (5G) mobile networks and the forthcoming sixth-generation (6G) will bring about fundamental changes in the way we communicate, access services and entertainment. In the context of security, inarguably, 5G security enhancements present a big improvement with respect to LTE. However, as the complexity of the application scenarios increases with the introduction of novel use cases, notably ultra-reliable low latency (URLLC), massive machine type communications (mMTC) and autonomous cyberphysical systems (drones, autonomous cars, robots, etc.), novel security challenges arise that might be difficult to address using the standard paradigm of complexity based classical cryptographic solutions. 

Specific use cases with open security issues are described in detail in a number
of 3GPP technical reports, e.g., on the false base station attack scenario \cite{false_BS} and on the security issues in URLLC \cite{URLLC_security}. Indeed, for beyond 5G (B5G) systems, there exist security aspects that can be further enhanced by exploiting different approaches, as classical mechanisms either fall short in guaranteeing all the security and privacy relevant aspects, or, can be strengthened with mechanisms that could provide a second layer of protection.

In the past years, physical layer security (PLS) \cite{bloch_barros_2011} has been studied and indicated as a possible way to emancipate networks from classical, complexity based, security approaches. Multiple white papers on the vision for 6G incorporate physical layer security, e.g., \cite{Oulu_white-paper, NOKIA_6G_security_trust, Hexa-X}, as well as in the IEEE International Network Generations Roadmap (INGR) 1st and 2nd Editions \cite{ieee_roadmap}. Motivated by the above, a key point of this paper is to showcase how PLS and in general security controls at the PHY level can be exploited towards securing future networks. 

One of the most promising and mature PLS technologies concern the distillation of symmetric keys from shared randomness, typically in the form of wireless fading coefficients. Within the channel's coherence time, small scale fading  is reciprocal, time-varying and random in nature and therefore, offers a valid, inherently secure source for key agreement (KA) protocols between two communicating parties. This is pertinent to many forthcoming B5G applications that will require strong, but nevertheless, lightweight KA mechanisms, notably in the realm of Internet of things (IoT). 

With respect to authentication, there are multiple PLS possibilities, including physical unclonable functions (PUFs),  wireless fingerprinting and high precision localization. Combined with more classical approaches, these techniques could enhance authentication in demanding scenarios, including (but not limited to) device to device (D2D) and Industry 4.0. Note that according to the 6G vision, as a network of (sub)networks, authentication might be required independently for access to the local (sub)network and to the core network, making the adoption of RF and device fingerprints a viable alternative for fast authentication of local wireless connections.

In parallel, mmWave and subTHz bands require the use of a huge number of antennas and pencil sharp beamforming. Consequently, a viable scenario for the wiretap channel can be substantiated, without any assumptions regarding the hardware (number of antennas, noise figure, etc.) or the position of a potential eavesdropper. Similarly, visible light communications (VLC) systems offer respective use cases. It is therefore pertinent to discuss advancements in wiretap secrecy encoders. The interplay between secrecy and privacy in finite blocklengths is another aspect that emerged from recent fundamental results in finite blocklength secrecy coding and should be highlighted.

Furthermore, new types of attacks have to be accounted for. In particular, there is mounting concern for potential jamming attacks and pilot contamination attacks during beam allocation and entry phases of nodes into the network \cite{Nokia_pilot_contamination}. Clearly, such attacks cannot be addressed with standard cryptographic tools and the required solutions can only emerge at the PHY, potentially in the form of jamming-resilient waveform and code design.

Finally, a less considered aspect relates to anomaly / intrusion detection by monitoring hardware metrics. This can be either used for distributed anomaly detection in low-end IoT networks, i.e., by monitoring memory usage, Tx and Rx time, debug interface of devices, or,  for more generalized anomaly detection of devices of untrusted manufacturers, etc. Such approaches could help lessen the monitoring overhead of centralized approaches and could provide new approaches towards the identification of the source of the anomaly~\cite{anomaly_detection1}. 

Looking at the bigger picture, future security controls will be adaptive and context-aware \cite{Context_aware_security_Chorti2022}. In this framework, rethinking the security design bottom up can provide low-cost alternatives. In particular, 
\begin{enumerate}
\item[1] PLS can provide information-theoretic security guarantees with lightweight
mechanisms (e.g., using LDPC, Polar codes, etc.); 
\item[2] Hybrid crypto-PLS protocols can provide fast, low-footprint and low-complexity solutions for  issues such as in \cite{false_BS} and \cite{URLLC_security};
\item[3] PLS can act as an extra security layer, complementing other approaches, enhancing the trustworthiness of the radio access network (RAN);
\item[4] PLS is inherently adaptive and can leverage the context and the semantics of the data exchanged.

\end{enumerate}

%Our motivation in this tutorial on PLS stems from the fact that in B5G PLS emerges as a
%complementary means to enhance the security in demanding low latency and massive connectivity
%scenarios. A few supporting examples include: 1) the security vulnerabilities identified in [1] arise during
%the establishment of the radio link; in this aspect, standard security protocols that build on the premise
%that the communication link has already been established, cannot offer solutions when this is not the
%case, whereas, PLS schemes can be seamlessly incorporated (e.g., can be interwoven with channel
%estimation); 2) in the realm of massive IoT in which standard authentication and key distribution
%/management becomes challenging (it is unrealistic to use digital certificates for billions of devices), PLS
%can offer complementary, device oriented solutions; 3) in 6G we will move away from the standard
%client/server networking paradigm on which the most successful security protocols build on and
%incorporate D2D and D2Edge at massive scales; and 4) the standard “rigid” on-or-off security approach of
%current protocols might not be the best fit to future generations of “semantic” communications between
%smart devices. A further benefit comes from the fact that PLS techniques – if implemented correctly –
%can offer quantum resistance. In this sense, PLS could pave the way out of the low latency impasse
%introducing novel lightweight mechanisms to post-quantum security.

In the following we will provide a comprehensive review of fundamental, cutting edge results in PLS and showcase how PLS can be employed to achieve many of the standard security goals, notably confidentiality, authentication, integrity. To this end, and, in order to provide a platform for a fair comparison to standard crypto schemes and a discussion on the potential advantages of hybrid PLS-crypto systems, we will first review fundamental cryptographic concepts and goals in Section \ref{sec:Crypto}. Next, Section \ref{sec:Motivation} gives a brief motivation on why PLS should be considered for the 6G. In Section \ref{sec:Wiretap} the wiretap channel theory will be presented (focusing on information theoretic characterizations for the finite blocklength) along with some recent results for privacy in sensing systems. Subsequently Section \ref{sec:SKG} discusses the topic of secret key generation (SKG) from shared randomness and highlights two subtle points concerning the pre-processing of the observation channel coefficients and coding methods in the short blocklength, furthermore, jamming attacks and countermeasures are discussed~\cite{Mitev_globecom_2019_MiM,jamming_countermeasure2}. In Section \ref{sec:Authentication} hardware based and statistical methods used in authentication will be visited, focusing on localization based authentication \cite{localization_authentication1, Mitev2022_access} and physical unclonable functions. 
% , along with anomaly / intrusion detection \cite{anomaly_detection1,anomaly_detection2} leveraging the PHY will be presented in \ref{sec:Availability}.
Finally, future directions and the authors' aspirations for security controls at all layers in 6G will be presented in Section \ref{sec:Prospects}.

%\par The total capture of these methods is that if a CP exists in the monitoring period, the residuals are expected to deviate from that of the residuals in the training period.

%%%%%%%%%%%%%%%%%%%%%%%%%%%%%%%%%%%%%%%%%%%%%%%%%%%%%%%%%%%%%%%%%%%%%%%%%%%%%%%%%
%%%%%%%%%%%%%%%%%%%%%%%%%%%%%%%%%%%%%%%%%%%%%%%%%%%%%%%%%%%%%%%%%%%%%%%%%%%%%%%%%
\section{Background Concepts in Cryptography and Network Security}\label{sec:Crypto}
%%%%%%%%%%%%%%%%%%%%%%%%%%%%%%%%%%%%%%%%%%%%%%%%%%%%%%%%%%%%%%%%%%%%%%%%%%%%%%%%%
%%%%%%%%%%%%%%%%%%%%%%%%%%%%%%%%%%%%%%%%%%%%%%%%%%%%%%%%%%%%%%%%%%%%%%%%%%%%%%%%%

Starting with some fundamental concepts in cryptography, we will address questions that arise in the systematic study of any system. In particular, we will provide answers to the following questions: "what do we want to achieve?"; "what is the system model?"; "what are the underlying assumptions, and what are the desirable properties?" 

With respect to what we aim to achieve, typically any security system aims at reaching one or multiple of four fundamental goals. The first goal is to be able to provide data confidentiality, i.e., security against eavesdropping (passive attackers). The corresponding threat model involves two legitimate parties communicating in the presence of an eavesdropper. Typically, with the aid of encryption, confidentiality is ensured against passive attackers. 
The second major goal is that of data integrity, i.e., providing guarantees that as the data traverses through the network,  any modification or alteration of a message will be perceptible at the destination. The corresponding threat model involves an active attacker that in addition to intercepting messages also performs modifications. 
The third major security goal is authentication (user or device), while access control is a closely related topic. The threat model involves again an active attacker that potentially attempts to gain unauthorized access. 
% Modern asymmetric key % and there are some aspects of modern cryptography that provide very desirable properties, such as non-repudiation. We will not be discussing this here today, but just to mention it for completeness. 
Finally, the fourth goal is that of availability, i.e., users should not be denied services. The network should be resilient to active attacks that fall in the general category of ``denial of service''. 

With respect to the system model, as noted above, the basic system setting includes three nodes. Two legitimate parties, that are referred here Alice and Bob and an adversarial node that is typically referred to as Eve (passive eavesdropper) or Mallory (active attacker, i.e., man-in-the-middle). To securely transmit a message (plaintext) to Bob, Alice uses a secret key to first encrypt it to a ciphertext. The ciphertext is then propagated through the transmission medium and received at Bob. Bob can decrypt the ciphertext by using the same or a different type of key, depending on the underlying algorithm.

\subsection{Confidentiality}
% \textcolor{red}{Maybe we should shorten a bit this subsection? It is much bigger than the other 3. Or just move the first 3-4 paragraphs just before the subsection?}
To perform the operations above, i.e., encryption / decryption, Alice and Bob rely on the use of ciphers. A key feature of modern block ciphers is to exploit highly non-linear operations to induce confusion, i.e., to render statistical inference attacks impossible. A textbook example of a linear cipher that is badly broken is the substitution cipher in which each letter of the alphabet is moved $k$ positions to the right (or to the left), with $k$ changing per letter. Considering the English alphabet, this results in $25!$ possible key combinations, making a brute force attack impractical. However, due to the linearity of the operations (permutations), a frequency analysis of a (long enough) ciphertext suffices to guess the plaintext. 

A revolutionizing result in security was presented by Shannon in  $1949$ \cite{Shannon_1949}, when he demonstrated that perfect secrecy can be achieved if and only if (iff) the entropy of the secret key is greater or equal to the entropy of the plaintext. The corresponding scheme, known as one-time-pad, is implemented by xor-ing the plaintext with the key. Unfortunately, to perform the above, the key size must be at least equal to that of the data which raises the problem of key distribution. 

While one-time pad is impractical, it provided insight into how secrecy can be achieved. In particular, it inspired the family of stream ciphers that rely on the idea of inflating short key sequences to psedorandom sequences of the same size as the plaintext and xor-ing them. This is achieved through the use of pseudorandom number generators (PRNGs). Although they cannot provide perfect secrecy (entropy cannot increase by data processing as a consequence of the data processing inequality), their usage led to the introduction of a more practical concept, i.e., semantic security.   

The definition of semantic security for PRGNs relies on the indistinguishability between  their output and the output of a truly random source. More generally, semantic security ensures that a non-negligible statistical advantage cannot be accumulated by an adversary in polynomial time. For all practical purposes,  if a statistical advantage happens with probability higher that $2^{-30}$, e.g., one bit is leaked in one gigabyte of data, the system is considered broken (not semantically secure). %With respect to PRNGs, they are deemed as semantically secure if they can resist next bit predictors. 

A canonical example of modern block ciphers is the advanced encryption standard (AES). AES is a semantically secure symmetric block cipher which takes a $n$-bit plaintext ($n=128$) and a $k$-bit key ($k$ chosen from $128$, $192$, or $256$ bits, with AES-$256$ considered to be quantum resistant) as input and outputs a $n$-bit ciphertext. AES relies on a set of substitution and permutation operations including the use of substitution (S) boxes. A well structured S-box removes the relation and dependency between bits, making a (linear or differential) cryptanalysis attack impossible. To allow the re-use of a single key for multiple blocks, nonces can be used. Nonces are deterministic (e.g., a counter) or random (initialization vectors), chosen such that a pair (key, nonce) never repeats. The important message here is that, today's cryptographic mechanisms allow the use of a short key sequence (e.g., 96 Bytes of key material in TLS v1.3) for the encryption of very long data sequences (in the order of GBs), allowing to overcome the key issue with one-time pad.
 
\subsection{Data integrity}

Data integrity is achieved with message authentication codes (MACs). The principle of MACs is to append a small label (tag) to each message, which validates its integrity. A MAC consists of two algorithms: signing and verification. Similarly to confidentiality schemes, there are historical examples of broken integrity algorithms in which linear functions (e.g., cyclic redundancy checks) have been used to generate MACs. Modern signing algorithms (tag generation) leverage the use of secret keys and symmetric block ciphers to generate a $t$-bit tag for a $n$-bit message, with $t<<n$. %The tag is generated by inputting all blocks of the message in single iteration of the cipher, as opposed to confidentiality schemes where blocks are fed separately. 
Upon reception, the verification algorithm uses the key, the received message and the tag and outputs a binary decision, i.e., the integrity check is either successful or not. 

Building on the above, a naturally arising concept is the one of authenticated encryption (AE) which combines both confidentiality and integrity. Various options exist on how to perform the two operations. One approach, that is always correct and provably secure, is the so called encrypt-then-sign, i.e., after a plaintext is encrypted a tag is generated over the ciphertext. The receiver would  first check the integrity and iff successful would continue with decryption.

\subsection{Authentication}

The process of authentication relies on digital signatures, which in turn, are used to produce digital certificates. Digital certificate is data signed by a trusted third party (certificate authority (CA))  that ensures the authenticity of the its owner. A certificate contains information about the CA, the owner of the certificate, the validity of the certificate, etc. As an example, when a user accesses a public server, the server proves its authenticity by presenting a certificate signed from a CA. To achieve mutual authentication the user must enter a password information, provide biometric data, etc. %Once authentication is achieved, the two parties perform a key exchange session where a shared symmetric key is obtained. Once this is established, both enter a secure communication stage where data is encrypted and exchanged using AE schemes, as described above. 

% \subsection{Availability}
% \textcolor{blue}{This is a new section as we were missing it.}
% \textcolor{red}{As noted above, the fourth security goal is to ensure that data is available to users when needed. To limit the availability (e.g., block authentication procedures or de-synchronize connections) an attacker can launch denial of service attack. Unfortunately, there are no standardized solutions to prevent such attacks, instead, research is mostly focused on detection techniques. There are different type of machine learning and statistical methods that has been used to detect and redirect malicious traffic~\cite{SDN_security}. The main challenge is for networks with constant and high traffic flow where scanning all packets is impossible. }
%%%%%%%%%%%%%%%%%%%%%%%%%%%%%%%%%%%%%%%%%%%%%%%%%%%%%%%%%%%%%%%%%%%%%%%%%%%%%%%%%%%%%%%%
%%%%%%%%%%%%%%%%%%%%%%%%%%%%%%%%%%%%%%%%%%%%%%%%%%%%%%%%%%%%%%%%%%%%%%%%%%%%%%%%%%%%%%%%
\section{Motivation for Considering Physical Layer Security} \label{sec:Motivation}
%%%%%%%%%%%%%%%%%%%%%%%%%%%%%%%%%%%%%%%%%%%%%%%%%%%%%%%%%%%%%%%%%%%%%%%%%%%%%%%%%%%%%%%%
%%%%%%%%%%%%%%%%%%%%%%%%%%%%%%%%%%%%%%%%%%%%%%%%%%%%%%%%%%%%%%%%%%%%%%%%%%%%%%%%%%%%%%%%

Given the fact that all schemes discussed in the previous section are widely deployed and trusted, one question remains: What is the motivation in considering PLS?

PLS technologies can offer multiple security techniques: i) secrecy encoders for wiretap channels, ii) privacy preserving transmission, iii) secret key generation from shared randomness iv) physical unclonable functions for device authentication, and v) localization or RF fingerprinting based authentication. While crypto solutions can provide these functionalities for current standards, they face number of challenges when considering new and emerging technologies. First, latency requirements are getting more stringent than ever, bringing the need for faster authentication and integrity checks. Second, large scale IoT deployment requires flexible and easily scalable security solutions that could simultaneously satisfy different security levels. A third element comes from the rise of quantum computing which opens the need for quantum secure algorithms. Finally, a fourth motivation comes from the new PHY infrastructures where the number of operations performed at the edge are expected to rise dramatically. Therefore, it is of utmost importance to separate the security of the core network from the one at the edge and introduce new faster and lightweight security algorithms. The statements above are complemented with the following list: 
\begin{enumerate}
    \item Regarding latency, 3GPP has recently noted that delays should be minimized in two directions, delays incurred by the communication and delays incurred due to computational overhead. A particular case where computational overhead of current standards do not comply with the requirements is security. As an example, it has been shown that the verification of a digital signature, in a vehicular networking scenario using a $400$ MHz processor, exceeds the tolerated delays and requires approximately $20$ ms~\cite{Teniou18}. Such results hint that a revolutionizing actions are needed in that direction.
    \item Next, deploying billions of IoT devices is not inconceivable anymore. In 2016, it has been demonstrated that a Mirai sized attack (e.g., $6\times10^5$ bots) is plausible. The attack has been demonstrated over simple machines, e.g. water heater, however, controlling $6\times10^5$ can instantly change the demand in the smart grid by $3$ GW, which is comparable to having an access to a nuclear plant. Examples like this raise a lot of questions on the security of the IoT. 
    \item In 2017, the NIST started the investigation on the topic of quantum resistance and post-quantum cryptography. However, as it stands now, the state of the art is based on using  longer keys and increased complexity. This makes the mechanisms heavier which contradicts with the need for low latency and low footprint. Hence, post-quantum innovations at the moment are not well aligned to the expectations towards 6G networks.
    \item Finally, new PHY and networking structures are being developed for the next generation of communication technologies. The central idea is to enhance the role of AI edge intelligence. This is a key component, that can enable the use of PLS in 6G. More details regarding this point will be discussed in Sec. \ref{sec:Prospects}.
\end{enumerate}

In the following sections it will be discussed how PLS technologies can be employed and some fundamental results in the area will be showed.
 
%%%%%%%%%%%%%%%%%%%%%%%%%%%%%%%%%%%%%%%%%%%%%%%%%%%%%%%%%%%%%%%%%%%%%%%%%%%%%%%%%
%%%%%%%%%%%%%%%%%%%%%%%%%%%%%%%%%%%%%%%%%%%%%%%%%%%%%%%%%%%%%%%%%%%%%%%%%%%%%%%%%
\section{Confidentiality and Privacy Using PLS} \label{sec:Wiretap}
%%%%%%%%%%%%%%%%%%%%%%%%%%%%%%%%%%%%%%%%%%%%%%%%%%%%%%%%%%%%%%%%%%%%%%%%%%%%%%%%%
%%%%%%%%%%%%%%%%%%%%%%%%%%%%%%%%%%%%%%%%%%%%%%%%%%%%%%%%%%%%%%%%%%%%%%%%%%%%%%%%%

\subsection{Confidential transmission}

In this section two aspects of physical layer security will be discussed, i.e., data confidentiality and data privacy. In detail, the information theoretic formulations of these problems will be investigated. 

As noted in Section \ref{sec:Introduction} secure data transmission tends to be a higher layer issue, e.g., enabled by encryption. However, confidential data transmission becomes difficult when considering massive numbers of low cost and low complexity devices. This is where physical layer security can play an important role. The idea is, instead of having reliability encoding, i.e., error control coding separated from the encryption, we can use joint encoding schemes that provide both reliability and security.

This approach, known as wiretap coding, was proposed approximately half a century ago by A. Wyner~\cite{Wyner_wiretap_1975}. Wyner looked at a three terminal wireless channel, i.e., two legitimate users Alice and Bob, and an eavesdropper, Eve. He recognized that the channels between the terminals are not perfect, i.e., their transmission will be impacted by noise. Therefore, when Alice transmits, Bob and Eve will not see exactly what has been transmitted. Moreover, Bob and Eve will have different received signals as they have different noisy channels. Wyner was interested in whether Alice could send a message reliably to Bob, while keeping it secret from Eve. To answer, he looked at the reliable rate to Bob, versus the equivocation at Eve (conditional entropy of the message at Eve's receiver). Note that, perfect secrecy can be achieved if the reliable rate at which data is being transmitted to Bob equals to the equivocation of Eve. To measure these quantities Wyner introduced a new metric, named secrecy capacity, which is the maximum reliable rate that equals the equivocation. He further showed that, achieving positive secrecy capacity is possible, hence, confidential transmission can be performed without the use of secret keys. However, achieving positive secrecy capacity is possible iff, the measurements at Eve are degraded with respect to those at Bob. A plausible example is when the signal to noise ratio (SNR) at Bob is higher than the SNR at Eve. 

% So now if we zoom forward to the present, or maybe 10 years ago and through the present, there's been a resurgence of interest in these kinds of ideas. And basically, two things we learned from Shannon and Wyner are that we can transmit in secrecy, a perfect secrecy, using information theoretic criteria. But to do so, the legitimate receiver needs some kind of advantage over the eavesdropper. Either a shared secret like in Shannon's model or a better channel, as in Wyner's model. 

Now, thinking about the physical layer, it is clear that the properties of radio propagation, i.e., diffusion and superposition, provide opportunities to achieve positive secrecy capacity. For example, by using the natural degradeness over time (e.g., fading), by introducing an artificial degradeness to the eavesdropper (e.g., interference and jamming), or, by leveraging spatial diversity (e.g., multiple antenna systems and relays can create secrecy degrees of freedom).

\begin{figure}[!t]
    \centering
    \includegraphics[width=0.48\textwidth]{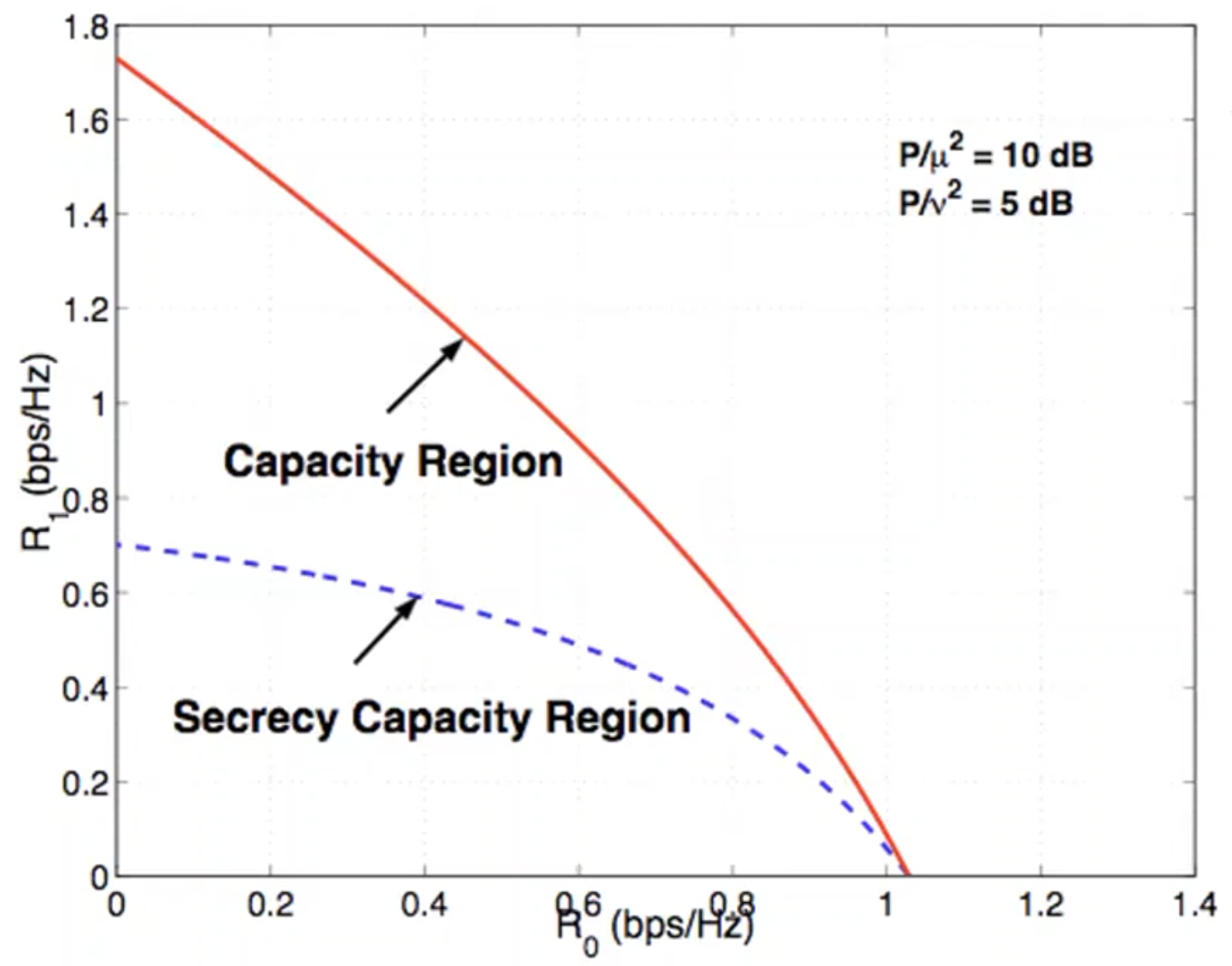}
    \caption{Achievable rates for the  Gaussian broadcast channel.}
    \label{fig:AWGN_SR}
\end{figure}

Based on the above, over the last fifteen years the idea of wiretap coding has been further examined considering several fundamental channel models: broadcast channel (one transmitter, multiple receivers), multiple access channel (multiple transmitters, one receiver), interference channels (multiple transmitters, multiple receivers); see e.g.~\cite{Poor_WirelessPLS_2017}. To illustrate the main results in the area, this work focuses on the broadcast channel~\cite{Poor_SecureFading_2008}. First, consider a Gaussian broadcast channel with Alice being a transmitter and Bob and Eve receivers. Assume two messages are transmitted: $M_1$ intended for both receivers and $M_2$ a secret message that is intended only for Bob. To define the capacity region we consider a degraded channel at Eve. In particular, it is assumed that the SNR level at Bob equals $10$ dB, and the SNR at Eve is $5$ dB. This is illustrated in Figure \ref{fig:AWGN_SR} where the horizontal axis gives the range of possible rates for the common message $M_1$, and the vertical axis gives the range of possible rates for the secret message $M_2$. The capacity region without secrecy constraints is shown with red solid curve and the secrecy capacity is indicated by the dashed blue curve. It can be observed that, if secrecy is required, part of the available capacity must be sacrificed in order to confuse the eavesdropper for that message. It is important to note that the amount to be sacrificed depends upon choosing a codeword that  randomizes the message w.r.t. Eve, but allows Bob to successfully verify it.

\begin{figure}[!t]
    \centering
    \includegraphics[width=0.48\textwidth]{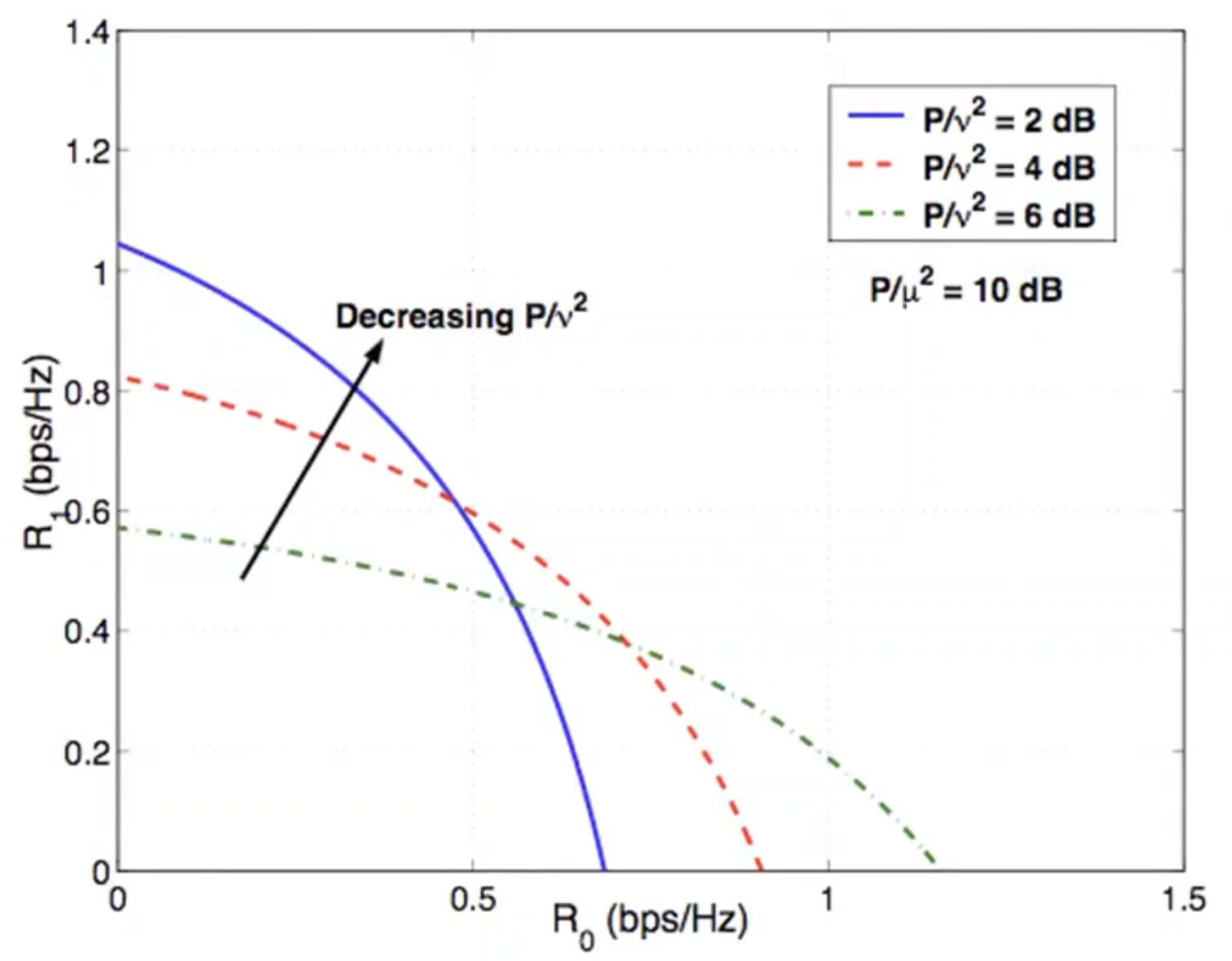}
    \caption{Achievable rates for the Gaussian broadcast channel considering variable SNR at Eve.}
    \label{fig:AWGN_SR2}
\end{figure}

Next, Figure \ref{fig:AWGN_SR2} shows the impact when the SNR at Eve varies. Similarly, the horizontal axis gives the common rate and the vertical axis gives the secrecy rate. The arrow shows that, if the SNR at Eve decreases, the range for the common rate shrinks and the range of secrecy rates increases. On the other hand, if the SNR at Eve reaches $10$ dB, the same level as Bob's SNR, the secrecy region  collapses. That is, if the second receiver is not degraded, secrecy rate becomes zero.

\begin{figure}[!t]
    \centering
    \includegraphics[width=0.48\textwidth]{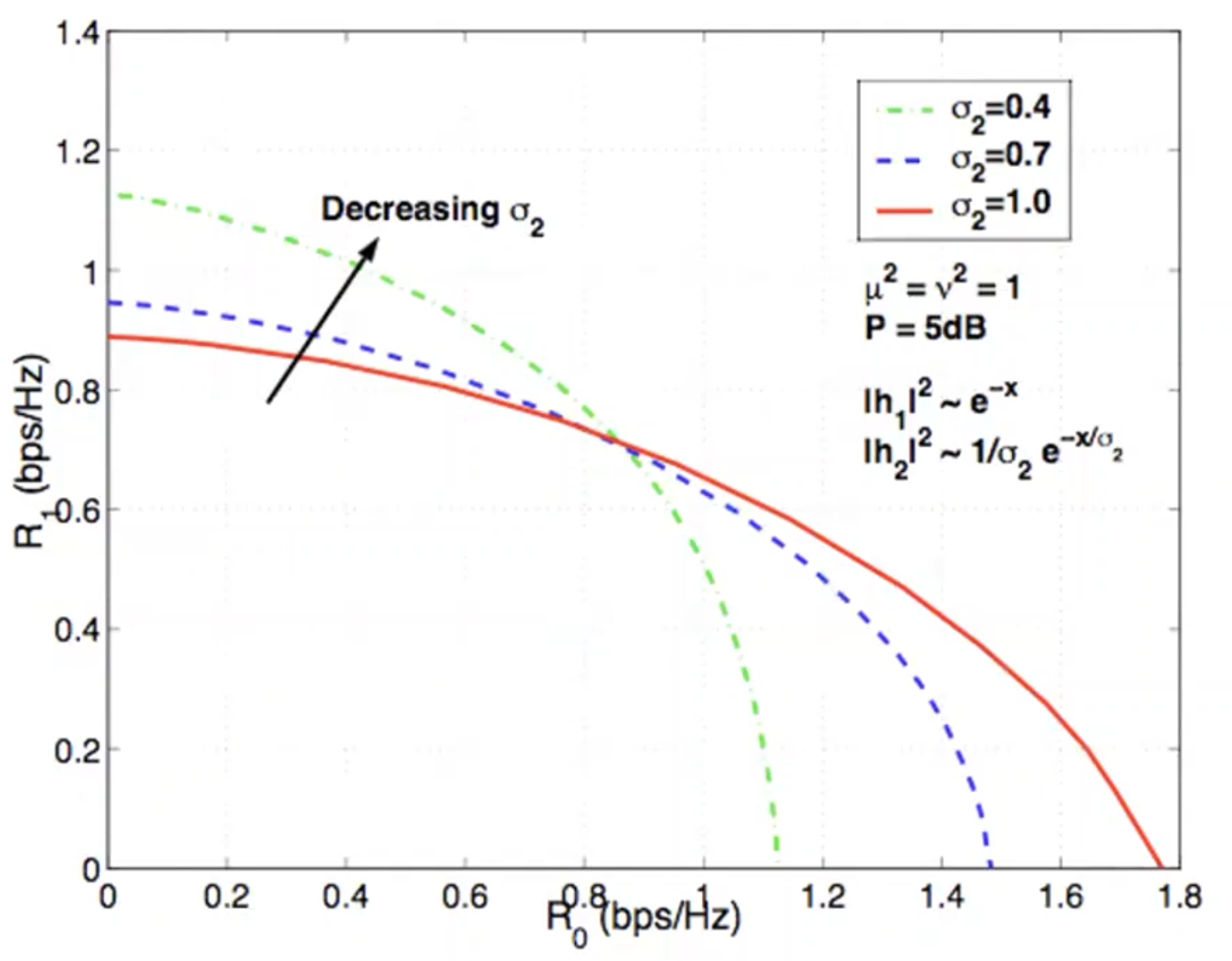}
    \caption{Achievable rates for the Rayleigh fading broadcast channel considering variable $\sigma^2$ at Eve. (From \cite{Poor_SecureFading_2008}.).}
    \label{fig:RAY_SR}
\end{figure}

Interestingly, things change when looking at a fading Gaussian broadcast channel. To illustrate this scenario we consider the same model, i.e., one transmitter, two receivers, one common message, and one secret message, but we assume that both the receivers have the same level of Gaussian noise, i.e., Bob and Eve have $5$ dB SNR. This is given in Figure \ref{fig:RAY_SR}. The difference between Bob and Eve is the fading parameter, i.e., Bob's experiences Rayleigh fading with a unit parameter, and Eve has Rayleigh fading with parameter $\sigma_2$. Note, a smaller $\sigma_2$, results in more intense fading. As before, when Eve's channel gets worse, i.e., $\sigma_2$ decreases, it can be seen that the range of common rates on the horizontal axis shrinks and the range of secret rates on the vertical axis increases. However, a distinction here is that if the two receivers observe the statistically identical channels (this is the case when $\sigma_2=1$), the secrecy capacity does not collapse as in the case of the Gaussian channel. This result holds under the assumption of perfect channel knowledge and follows from the fact that fading provides additional degrees of freedom leading to advantage during the time when other receivers experience deeper fade.

% Now, you might think about this model, for example, as being a model for a content distribution system where the broadcast message is sort of the generic content that everyone in the network is supposed to receive. And the secret message is really just some kind of premium content that only those nodes that have paid for the premium content are supposed to receive. So the fact that Eve is called an eavesdropper here doesn't necessarily mean that that terminal is a malicious terminal. It could just be a terminal that's not supposed to receive some of the content for economic reasons or what have you. 

A major issue concerning the results above comes from an information theoretic perspective. In particular, they are based on the assumption of infinite coding blocklength. Hence, it concerns the following scenario. Assume that a message $W$, that is encoded into a length-$n$ codeword, is transmitted into the channel. After passing through the wireless medium  noisy instances of the codeword are obtained by Bob and Eve. These codewords are then fed into Bob's and Eve's decoders. The desired property for this scenario is that for Bob to be able to reconstruct the codeword perfectly while at the same time, the leakage of the codeword to Eve is bounded by the quantity  $\delta$. In the original formulation by Wyner, the considered blocklength is infinity, i.e, $n$, the number of channel uses, is infinity. When $n \rightarrow \infty$, the probability of error at Bob, i.e., probability that he decodes to a $\hat{W}$ which is different compared to $W$ goes to zero. Additionally, the information leakage $\delta$ also goes to zero. The secrecy capacity for this case has been formulated as the difference between the mutual information between Alice, $X_A$, and Bob, $X_B$, and the mutual information between Alice and Eve, $X_E$, when considering the maximum from the channel input distribution $P_X$, i.e.,:
\begin{equation}
    C_S = \underset{P_X}{\max} \{I(X_A;X_B)-I(X_A;X_E) \}.
\end{equation}

This is an intuitive result, i.e., achieving positive secrecy capacity relies on the degradation of Eve's channel. The limitation of this theory is that it gives only  asymptotic results that are not suitable for low latency applications, such as in an IoT scenario. This opens the question: What is achievable in the non-asymptotic case?, and the answer depends on the finite blocklength information theory. Assume we have a source $W$, which can take $1,2,\dots, M$ possible values, i.e., it has $\log_2 M$ bits. The source is mapped using an encoder to a sequence, $X^n$, which is then passed through a channel. Due to noise, the receiver will observe a corrupted version of the transmission, i.e., $Y^n$, which is then decoded to $\hat{W}$. If the errors between $\hat{W}$ and $W$ are less than a particular value, $\epsilon$, the decoder could reconstruct the original source. In systems like this, the design of $nM\epsilon$ codes is of particular interest: $M$ the number of source symbols, $n$ the number of channel uses, and $\epsilon$ the upper bound on the reconstruction fidelity of the source at the output of the decoder. The fundamental limit for such a system is defined by the maximum $M$, i.e., the largest possible number of source symbols that can be transmitted through the channel in $n$ channel uses and be reconstructed at the decoder with error probability $\leq \epsilon$. Note that, $\underset{n\rightarrow \infty}{\lim}\frac{1}{n}\log_2(M)$ gives the Shannon's capacity where $\epsilon \rightarrow 0$. However, in an actual system $n$ and $\epsilon$ are finite values. Considering this, an approximation for $M^*$ was derived in \cite{Polyanskiy2010}, and it is given as
\begin{equation}
    \log M^*(n,\epsilon) = nC - \sqrt{n} V Q^{-1}(\epsilon) + \mathcal{O}(\log n), \label{eq:M_star}
\end{equation}
where $C$ gives the Shannon's capacity, $Q^{-1}(\epsilon)$ defines the tail of a standard Gaussian distribution evaluated at $\epsilon$, and $V$ is the channel  dispersion, which is the variance of the information density (note that Shannon's capacity is the mean of the information density).

% Now you may remember from your information theory that capacity is the expected value of the so-called information density evaluated at a random variable, x star, that has the optimal input distribution, the capacity achieving input distribution. Y star is the corresponding output distribution of the channel, so that's a random variable with expected value as the capacity. And then if you take the variance of that expected value, you get the dispersion. So it's quite intuitive that the second order term in the optimal rate here is the second moment of the quantity of which the leading term is the expected value. 

\begin{figure}[!t]
    \centering
    \includegraphics[width=0.48\textwidth]{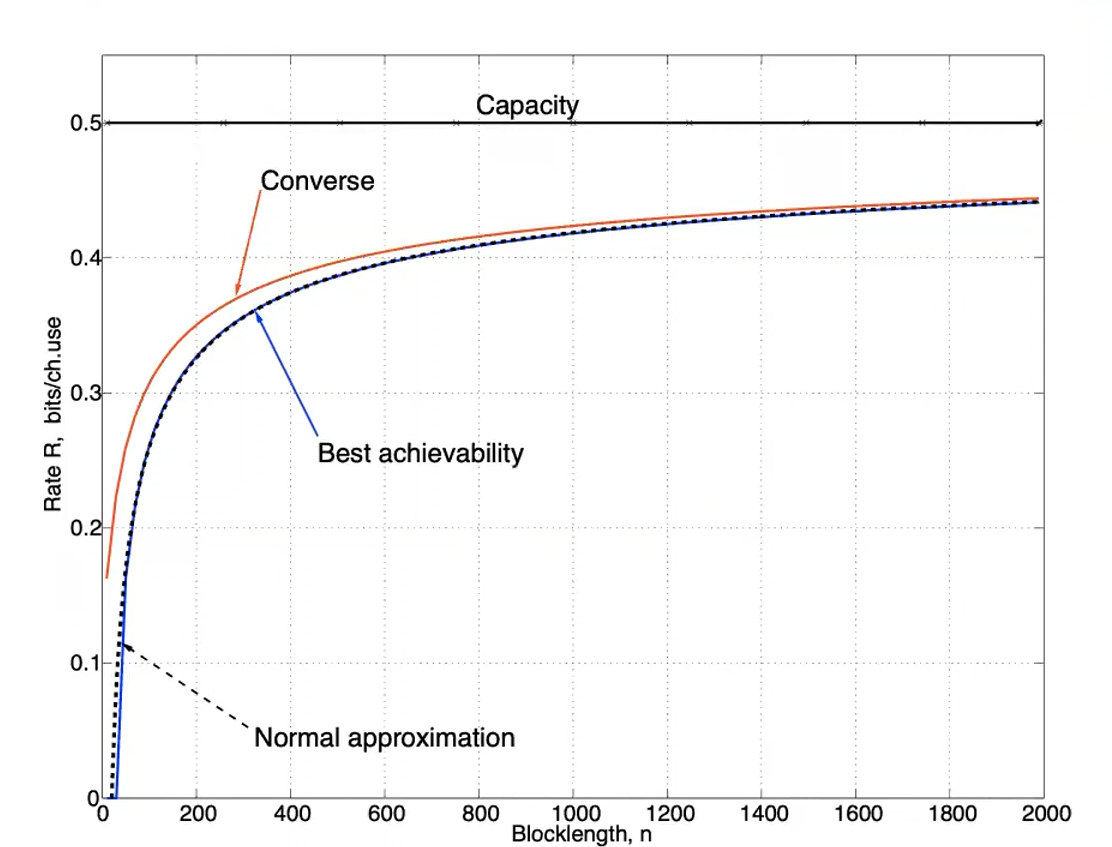}
    \caption{Upper and lower bounds on the capacity regions for short block length communication. SNR is equal to $0$ dB and $\epsilon=10^{-3}$. (From \cite{Polyanskiy2010}) }
    \label{fig:M_star-Polyanskiy2010}
\end{figure}

The result from Equation \eqref{eq:M_star} is illustrated in Figure \ref{fig:M_star-Polyanskiy2010}, where an AWGN channel is assumed with SNR equal to $0$ dB, $\epsilon=10^{-3}$ and $C=1/2$. The figure shows the upper bound and lower bound for the capacity for finite block lengths, denoted here by ``Converse'', and ``Best achievability'', respectively. Hence, the actual capacity, which remains to be found, lies between those two curves. While the gap between the curves is small for high values of $n$, it can be observed that for small values of $n$ the gap remains large, hence, further work in the area is required to obtain a more precise solution.

\begin{figure}[!t]
    \centering
    \includegraphics[width=0.48\textwidth]{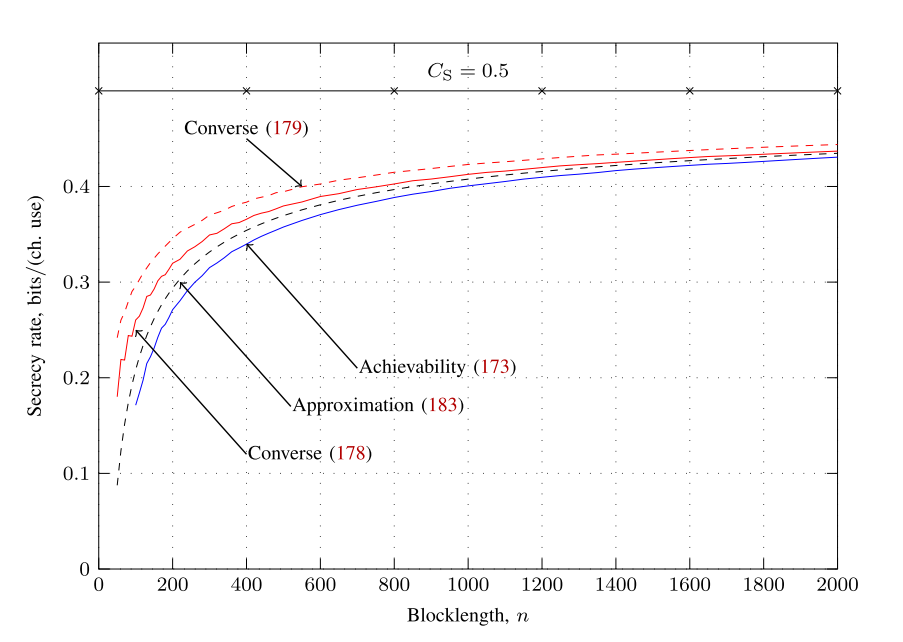}
    \caption{Upper and lower bounds on the secrecy capacity for short block length communication in a binary symmetric channel with crossover probability $p = 0.11$ and  $\delta=\epsilon=10^{-3}$. (From \cite{Yang2017}.)}
    \label{fig:R_star_semi-deterministic}
\end{figure}

Following the result for channel capacity, it has been just recently shown that the secrecy capacity in the finite blocklength scenario can also be approximated~\cite{Yang2017}. Fixing the error probability at Bob, $\epsilon$, the leakage at Eve, $\delta$, and the block length $n$, an approximation for the secrecy capacity is given as
\begin{equation}
    R^*(n,\epsilon,\delta) = C_S - \sqrt{\frac{V}{n}}Q^{-1}\left(\frac{\delta}{1-\epsilon}\right) + \mathcal{O}\left(\frac{\log n}{n}\right), \label{eq:R_star}
\end{equation}
where $V$ is defined similarly to the channel dispersion of \eqref{eq:M_star}. The result from Equation \eqref{eq:R_star} is illustrated in Figure \ref{fig:R_star_semi-deterministic}. The figure considers a binary symmetric wiretap channel with crossover probability $p = 0.11$,  $\delta=\epsilon=10^{-3}$ and $C_S=1/2$. A similar trend is observed as in the previous figure, the gap between upper bound (Converse) and lower bound (Best achievability) shrinks and widens as $n$ gets larger or smaller, respectively.

\begin{figure}[!t]
    \centering
    \includegraphics[width=0.48\textwidth]{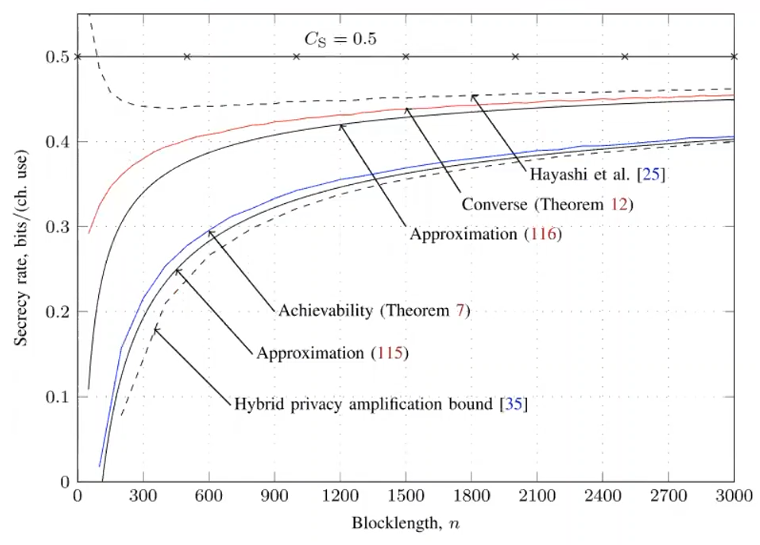}
    \caption{Upper and lower bounds on the secrecy capacity for short block length communication in a Gaussian wiretap channel. SNR at Bob is equal to $3$ dB and SNR at Eve equals to $-3$ dB. (From \cite{Yang2017}.)}
    \label{fig:R_star_gaussian}
\end{figure}

This has also been evaluated for a Gaussian wiretap channel and the result is illustrated in Figure \ref{fig:R_star_gaussian}. The SNR at Bob here equals $3$ dB, and the SNR at Eve  equals $-3$ dB. It can be observed that the gap between achievability and converse is even larger for this scenario. However, what is important to mention here is that the upper bound, when considering finite block lengths, is far from the asymptotic secrecy capacity, $C_S$. This shows that research on emerging IoT technologies should not rely on asymptotic results and should focus on the investigation of short block length communications. 

\subsection{Privacy in sensing systems}

Differently from secrecy, where the concern is about restricting a malicious party from getting access to the transmission, in the case of privacy, the goal is to keep part of the information secret from other parties, including the legitimate receiver (Bob). A simple way to ensure there is no privacy leakage is to deny access to Bob, however, without having a recipient the data source becomes useless. Therefore, it is important to study, which part of the data can be shared, such that the message is successfully and securely transmitted, while  the privacy leakage is minimized. 

This section focuses on the problem of privacy leakage with particular focus on sensing systems. Such systems include smart meters, cameras, motion sensors, i.e., devices that generate useful data for companies who provide users with particular service (alarm, power supply, etc.). While companies can use the data to improve their services, the full access to it endangers the privacy of users. 

\begin{figure}[!t]
    \centering
    \includegraphics[width=0.48\textwidth]{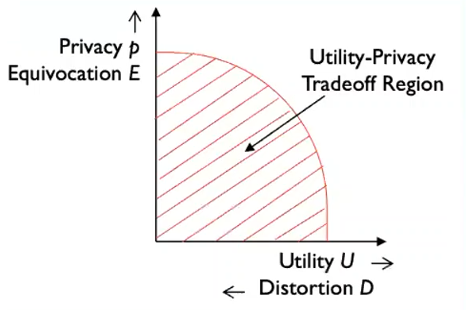}
    \caption{Trade-off between privacy and usefulness of data.}
    \label{fig:privacy-distortion}
\end{figure}

The above hints towards that, there is a fundamental trade-off between privacy and usefulness of data (distortion). This is illustrated in Figure \ref{fig:privacy-distortion}. If the data is completely private, i.e., its equivocation at Bob is high, the data becomes useless and it is fully distorted. Contrarily, if the data is fully accessible, i.e., it has low distortion at Bob, then its equivocation goes to zero and the data is not private.

\begin{figure}[!t]
    \centering
    \includegraphics[width=0.48\textwidth]{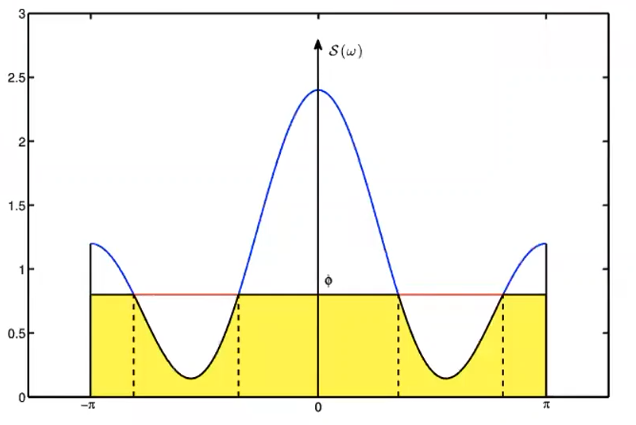}
    \caption{Privacy-utility trade-off characterized by a reverse water-filling.}
    \label{fig:hidden_Gauss-Markov}
\end{figure}

Now, when considering a specific application, i.e., smart meters, the trade-off can be specified as follows: a smart meter measures the electricity usage in almost real time, hence, having the utility of providing users with information on their usage, but in the same time it leaks this information to the power supply company who can use it to trace in-home activities~\cite{Poor_privacy-smartMeter_2018}. One way to model this problem is through a hidden Gauss-Markov model. This is given in Figure \ref{fig:hidden_Gauss-Markov} where the hidden state is the intermittent state, e.g., turning your toaster on, your kettle on, etc. The figure captures a smart meter trace, and shows that the privacy-utility trade-off for this model can be characterized by a reverse water-filling~\cite{Poor_SmartMeterPrivacy_2013}. The trade-off here is defined by the water level $\phi$, such that all signals with power lower than $\phi$ are being suppressed by the meter, while all signals above are being be transmitted (and leaked) by the meter. Therefore, the value of $\phi$ defines the amount of privacy that the user is willing to sacrifice to increase his utility.

\begin{figure}[!t]
    \centering
    \includegraphics[width=0.48\textwidth]{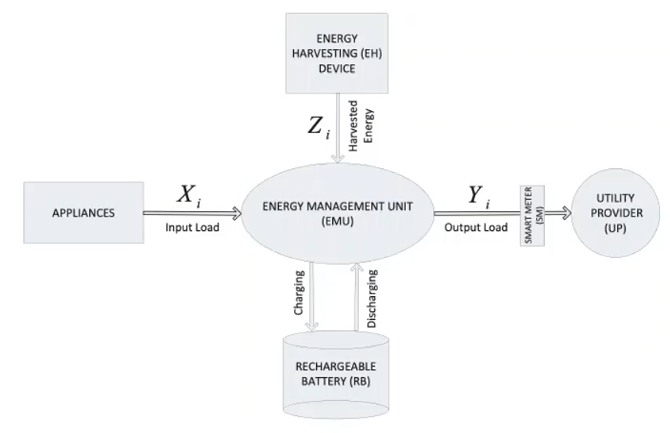}
    \caption{Privacy-utility trade-off characterized by a measuring wasted energy versus information leakage.}
    \label{fig:control_model}
\end{figure}

\begin{figure}[!t]
    \centering
    \includegraphics[width=0.48\textwidth]{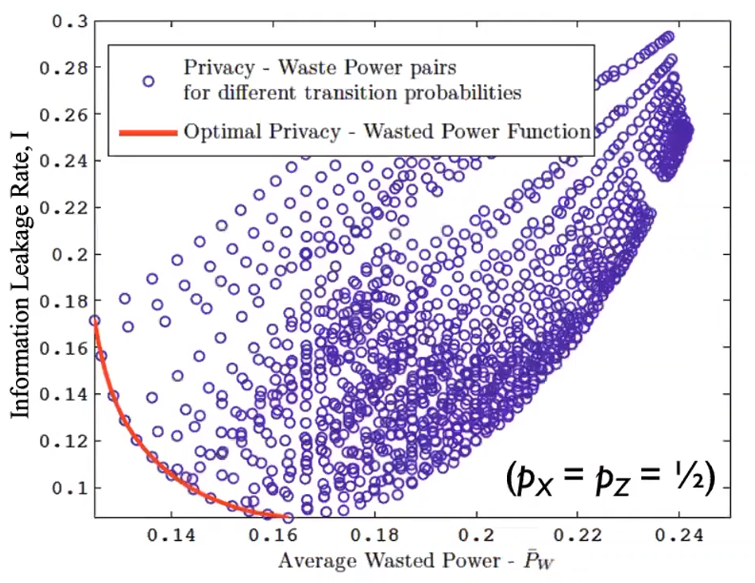}
    \caption{Wasted power versus information leakage when considering a control approach.}
    \label{fig:privacy-energy_binary-values}
\end{figure}

Another way to approach the same problem is through using control, i.e., actively controlling what the meter sees based on storage and energy harvesting~\cite{Poor_smartGrid_2020}. This is illustrated in Figure \ref{fig:control_model}, where the utility-privacy trade-off for this model is captured by measuring wasted energy versus information leakage. Presenting this control approach as a Markov model allows to numerically determine the efficient frontier. This is given in Figure \ref{fig:privacy-energy_binary-values}, where the red curve gives the optimal trade-off of wasted power versus information leakage. 

Another example is when considering the case of competitive privacy. In competitive privacy, there are multiple agents (Bobs) each having own privacy utility trade-offs. On one hand, there are multiple interacting agents who are competing with one another, but, on the other hand, the agents have coupled measurements. In detail, each agent wants to estimate its own parameters and can help other agents by sharing data but does not want to compromise his own privacy. 

% Well, just to give some examples, you might think about, well, we could think about a sensor network, which is a classical IoT type example. Where, say, two oil companies are looking for oil under the water. Well, they could each help each other by sharing measurements, right? But on the other hand, they're competitors. They don't want to give away their own measurements any more than they have to. So there's a trade-off that they can enact, but it's a competitive one. 

% You might also think about in the electricity markets, the electric energy providers share the same distribution system and transmission system, which they need to optimize on a daily basis. So that can be done by sharing data. On the other hand, they're competing entities in the marketplace. And so sharing data also compromises some privacy they have. So there's a competitive privacy there. 

% Also, in, say, in a warfare situation, you might have multiple allies who are allied against a common enemy. But they may not be the best friends. So they can certainly help each other by sharing data, sensing data, say, radar data. But on the other hand, they may be a little cautious about sharing all their data. So all of these are situations where you have competitive privacy. 

This competitive scenario can be represented as a linear measurement model~\cite{Poor_Belmega_2015}. Utility can be measured in terms of mean squared error on the state estimation and privacy can be measured in terms of information leakage. In fact, it has been shown that this reduces to a classical problem, known as the Wyner-Ziv problem or the distributed source coding problem. Particularly, it has not been discussed what is the optimal amount of information that must be exchanged, but it has been shown that the optimal way to exchange information is by using Wyner-Ziv coding. Next, depending on the scenario a simple way to find the optimal amount of information is through the use of game theory.

Finally, an important conclusion for this section is that information theory can help us understand the fundamental limits of security and privacy. While mainly theoretical constructs have been discussed, it is clear that there is a need to connect the theoretical analyses to real networks. Building on the above, some emerging research directions include finite blocklength analysis (short packet low latency communication), scaling laws for large networks (channel models that consider massive networks) and practical coding schemes. 

%%%%%%%%%%%%%%%%%%%%%%%%%%%%%%%%%%%%%%%%%%%%%%%%%%%%%%%%%%%%%%%%%%%%%%%%%%%%%%%%%
%%%%%%%%%%%%%%%%%%%%%%%%%%%%%%%%%%%%%%%%%%%%%%%%%%%%%%%%%%%%%%%%%%%%%%%%%%%%%%%%%
\section{Secret Key Generation Using PLS} \label{sec:SKG}
%%%%%%%%%%%%%%%%%%%%%%%%%%%%%%%%%%%%%%%%%%%%%%%%%%%%%%%%%%%%%%%%%%%%%%%%%%%%%%%%%
%%%%%%%%%%%%%%%%%%%%%%%%%%%%%%%%%%%%%%%%%%%%%%%%%%%%%%%%%%%%%%%%%%%%%%%%%%%%%%%%%
This section focuses on several aspects concerning SKG. First, it provides an overview on how to extract symmetric keys from shared randomness, then it shows how SKG can be incorporated in actual crypto systems, and finally, it discusses how the SKG process can be made resilient to active attacks.

%%%%%% CONTINUE HERE!!!!!!!!!!!!!!!
\subsection{Secret key generation}

Generally, the SKG protocol consists of three steps: advantage distillation, information reconciliation, and, privacy amplification. Assuming two legitimate parties, e.g., Alice and Bob, the steps can be summarized as follows. In the first step, Alice and Bob exchange pilot signals during the coherence time of the channel, and obtain correlated observations $Z_A$ and $Z_B$, respectively. In the second step, their observations are first quantized and then passed through a distributed source code type of decoder. During this step Alice (or Bob) shares side information, which is used by Bob to correct errors at the output of his decoder. Hence, at the end of this step both parties obtain a common binary sequence. Finally, to produce a maximum entropy key and suppress the leaked information, privacy amplification is performed. In this last step, Alice and Bob apply an irreversible compression function (e.g., hash function) over the reconciled bit sequence. This produces a uniform key that is unobservable by adversaries.

There are few important points that need to be taken into account for the success of the SKG process. First, channel measurements represent a mixture of large scale and small scale fading components. In multiple studies, it has been demonstrated that the large scale component is strongly dependent on the location and the distance between users, which makes it predictable for eavesdroppers. Therefore, to distill a secret key, Alice and Bob should either remove this part from their measurements and generate the key using the unpredictable small scale components or should compress more at the privacy amplification. This point is further discussed in Section \ref{sec:Prospects}. Second, the SKG protocol should follow all the steps described above, and no steps should be skipped. As an example, skipping the privacy amplification would give Alice and Bob longer key sequence, however, the key sequence is vulnerable to different attacks~\cite{PA}. Third, it is important that, Alice and Bob do not transmit information related to their observations, as this could be exposed to eavesdroppers in the vicinity. Forth, Alice and Bob should respect the coherence time and coherence bandwidth of the channel, such that their subsequent measurements are decorrelated in time and frequency. This allows them to generate random and unpredictable bit sequences.
% While there are machine learning techniques that can help in tackling this problem, this is still an open research topic.
Finally, as mentioned in the previous section, further testing of short blocklength encoders is necessary in order to identify the optimal solution for SKG.  

\begin{figure}[!t]
\centering
\includegraphics[clip, trim=0cm 0cm 0cm 0cm,width=0.48\textwidth]{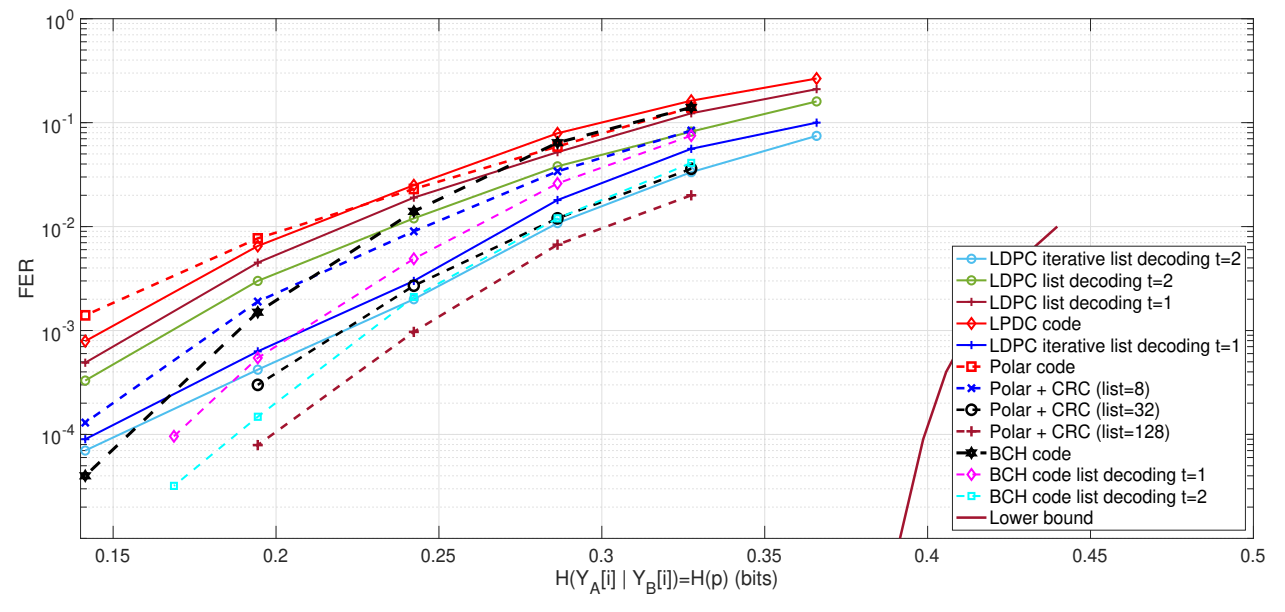}
\caption{  FER performance of reconciliation codes compared to the   lower bound from \cite{Tyagi2015} for $n = 128$. (From \cite{Mahdi_rec_codes2021}.)} \label{fig:128}
\end{figure}

\begin{figure}[!t]
\centering
\includegraphics[clip, trim=0cm 0cm 0cm 0cm, width=0.48\textwidth]{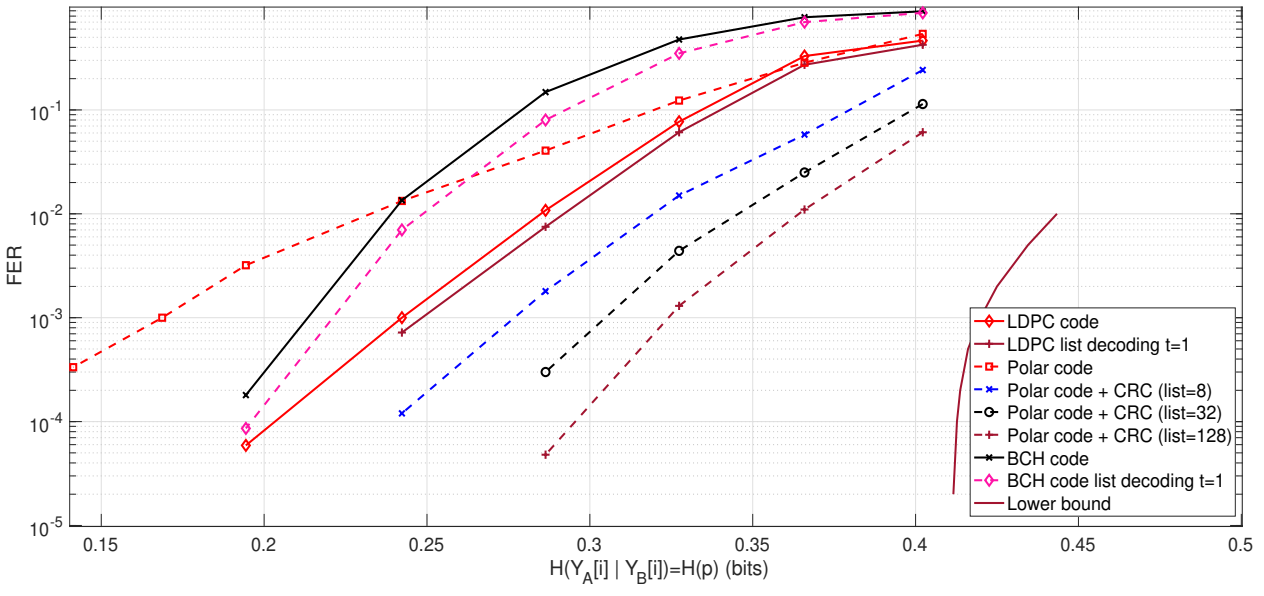}
\caption{FER performance of reconciliation codes compared to the  lower bound from \cite{Tyagi2015} for $n = 512$. (From \cite{Mahdi_rec_codes2021}.)} \label{fig:512}
\end{figure}

Regarding the last point, Figures \ref{fig:128} and \ref{fig:512} show a comparison between an upper bound, evaluated in \cite{Tyagi2015}, versus information reconciliation rates achieved using of LDPC, polar codes and BCH codes~\cite{Mahdi_rec_codes2021}. Both figures $n=128$ and $n=512$ show that polar codes with CRC and BCH codes with list decoding outperform the other approaches, making them good candidates for reconciliation decoding. Note that such type of encoders are already used in 5G for different purposes. 

%%%%%%%%%%%%%%%%%%%%%%%%%%%%%% continue here

%%%%%%%%%%%%%%%%%%%%%%%%%%%%%%%%%%%%%%%%%%%%%%%%%%%%%%%%%%%%%%%%%%%%%%%
%%%%%%%%%%%%%%%%%%%%%%%%%%%%%%%%%%%%%%%%%%%%%%%%%%%%%%%%%%%%%%%%%%%%%%%
\subsection{Secret key generation in hybrid crypto systems} \label{sec:hybrid_PLS}
%%%%%%%%%%%%%%%%%%%%%%%%%%%%%%%%%%%%%%%%%%%%%%%%%%%%%%%%%%%%%%%%%%%%%%%
%%%%%%%%%%%%%%%%%%%%%%%%%%%%%%%%%%%%%%%%%%%%%%%%%%%%%%%%%%%%%%%%%%%%%%%

Building on the above, we continue with a particular example on how SKG can be incorporated in hybrid security cryptographic schemes. In detail, it will be discussed how to build a SKG-based authenticated encryption. Three ingredients are needed to formulate this problem: 

\begin{itemize}
    \item A SKG scheme $\verb"G": \mathbb{C}\rightarrow \mathcal{K} \times \mathcal{S}$, that takes channel measurements as input and generates a key $\mathbf{k}$ and side information $\mathbf{s}$.
    
    \item A symmetric encryption algorithm, i.e., a pair of functions $\verb"Es": \mathcal{K}\times \mathcal{M} \rightarrow \mathcal{C_T} $ and $\verb"Ds": \mathcal{K}\times \mathcal{C_T} \rightarrow \mathcal{M}$, for encryption and decryption, respectively, where $\mathcal{C_T}$ defines the ciphertext space and $\mathcal{M}$ the message space.

    \item A message authentication code (MAC) algorithm, given as $\verb"Sign": \mathcal{K}\times \mathcal{M}\rightarrow \mathcal{T}$, for signing and $\verb"Ver": \mathcal{K}\times \mathcal{M} \times \mathcal{T} \rightarrow (yes, no)$, for verification, where $\mathcal{T}$ defines the tag space. 
\end{itemize}

Now, the components can be combined as follows:
\begin{enumerate}
    \item SKG is performed between Alice and Bob as:
     \begin{equation}
        \verb"G"(\mathbf{h})= \left(\mathbf{k}, \mathbf{s_A}\right),
    \end{equation}
    where $\mathbf{h}$ represents the channel measurements, $\mathbf{k}$ the generated key after privacy amplification and $\mathbf{s_A}$ is Alice's side information that has to be transmitted to Bob to finalize the process. 
    \item Before transmitting $\mathbf{s_A}$ to Bob, Alice breaks her key into two parts $\mathbf{k}=\{\mathbf{k}_e, \mathbf{k}_i\}$, generates a ciphertext as $\mathbf{c}=\verb"Es" (\mathbf{k}_e, \mathbf{m})$ and signs it as $\mathbf{t}=\verb"Sign"(\mathbf{k}_i, \mathbf{c})$. Afterwards she transmits to Bob the concatenation of $[\mathbf{s}_A||\mathbf{c}||\mathbf{t}]$, i.e., in a single message she can transmit the side information and her message.
    \item Upon receiving the above, Bob uses the side information $\mathbf{s}_A$, to finish the SKG process, i.e., to obtains the key $\mathbf{k}$. Then, he checks the integrity of the received ciphertext as $\verb"Ver" (\mathbf{k}_i, \mathbf{c}, \mathbf{t})$ and if successful he decrypts and obtain the message $\mathbf{m}$.
\end{enumerate}

Differently from the standard SKG scheme, where SKG is performed in parallel at both nodes and data exchange happens only after the key generation is finalized, in the scheme above Alice completes the SKG locally and then transmits in a single go the ciphertext, the tag, and, the side  information (e.g., syndrome). Then Bob uses the syndrome to complete the SKG and performs the authenticated decryption. This small change in the standard procedure shows how PLS can be easily combined with standard crypto schemes.

Such approaches bring new opportunities. For example, the scheme above opens the problem of transmission optimization. Consider a scenario with multiple subcarriers used for transmission. The subcarriers can then be split into two subsets, a subset $\mathcal{D}$ used for transmitting encrypted data and a subset $\bar{\mathcal{D}}$ used for transmitting side information (syndromes). This transmission scheme can be optimized considering several constraints. The first constraint comes from the world of cryptography, i.e., based on the choice of cryptographic cipher we can define the amount of data to be encrypted with a single key. This can be captured by the following constraint:
\begin{equation}
    C_{SKG}\geq \beta C_D, \quad 0 < \beta \leq 1, \label{eq:beta_constraint} 
\end{equation}
where $C_{SKG}$ defines the key generation rate, $C_D$ defines the data rate and $\beta$ is a quantity that relates the key size to the data size that will be encrypted, e.g., $\beta=1$ corresponds to a one-time pad cipher. The second constraint comes from the world of information theory. It relates the necessary (side information) syndrome rate $C_R$ and the SKG rate as follows:
\begin{equation}
    C_R\geq \kappa C_{SKG}, \label{eq:kappa}
\end{equation}
where $\kappa$ defines minimum number of reconciliation bits with respect to the key bits. It is a parameter defined by the type of the encoder/decoder used for SKG, e.g., for a $\frac{k}{n}$ block encoder $\kappa=\frac{n-k}{k}$.

Further constraints that can be incorporated are power constraint:
\begin{equation}
\sum_{j=1}^N p_j \leq NP,\;\; p_j \geq 0,\; \forall j \in\{1, \dots, N\},\label{eq:power}
\end{equation}
and a channel capacity constraint, i.e.,  
\begin{equation}
C_D+C_R\leq C \label{eq:C},
\end{equation}
where $N$ gives the number of subcarriers, $P$ is the power limit per subcarrier and $C$ is the total capacity of the channel. The objective of the problem can then be defined as: 
\begin{equation}
    \max_{p_j, j\in\mathcal{D}}C_D \quad \text{s.t. \eqref{eq:beta_constraint}, \eqref{eq:kappa}, \eqref{eq:power}, and \eqref{eq:C}}  \label{eq:optimisation}
\end{equation}

The problem can be turned into a combinatorial optimization problem which can be solved optimally using dynamic programming techniques or sub-optimally using heuristic approaches. Overall, this problem shows how physical layer aspects can be related to cryptographic schemes, in the form of a hybrid security scheme, and provide new opportunities for cross layer optimization. 

\begin{figure}[!t]
    \centering
    \includegraphics[clip, trim=3.5cm 9.4cm 4cm 10cm,width=0.48\textwidth]{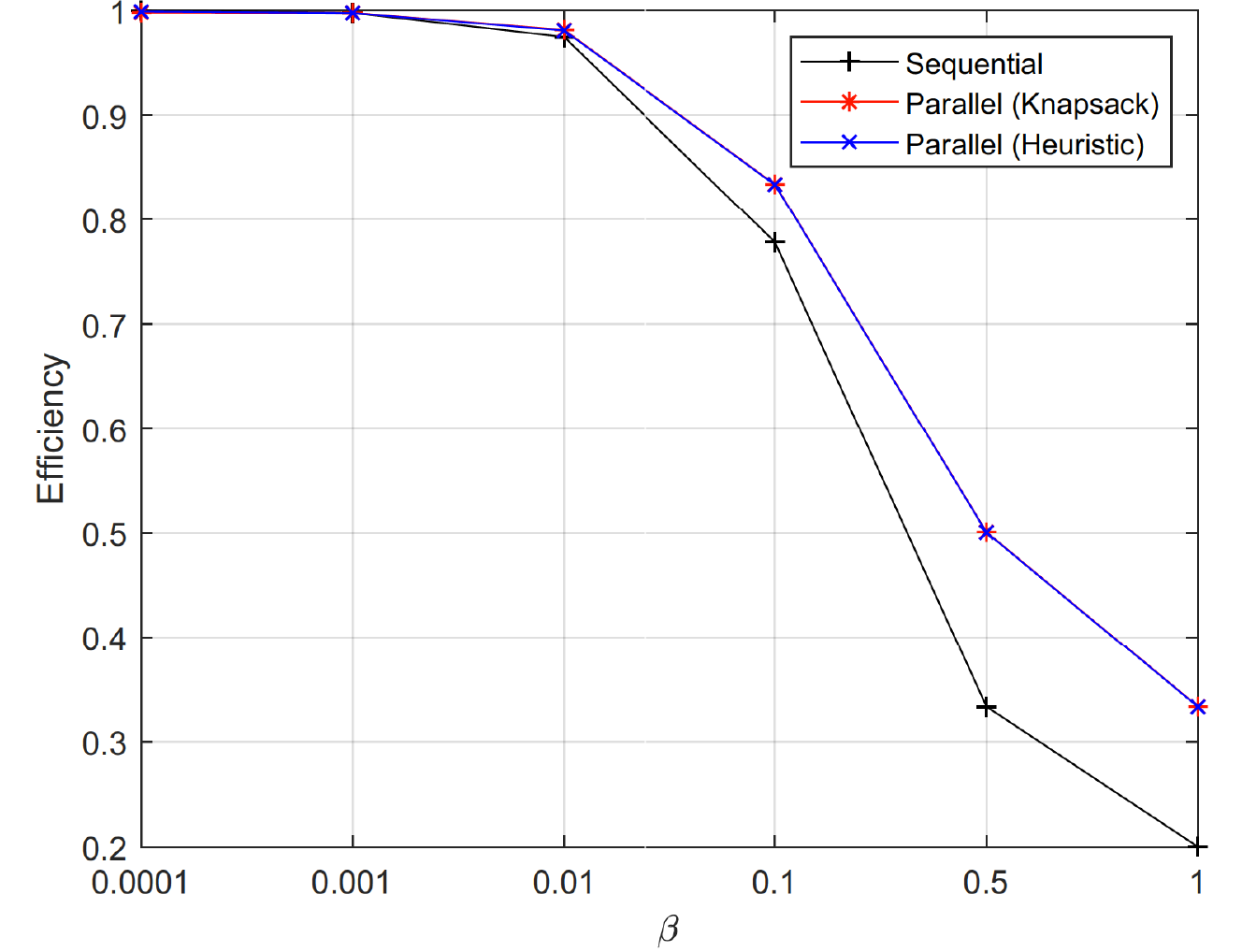}
    \caption{Efficiency comparison for $N = 64$, SNR$=10$ dB and $\kappa = 2$. (From \cite{Mitev_sub-scheduling_globecom2019}.)}
    \label{fig:parallel_sequential}
\end{figure}

The problem was solved in \cite{Mitev_sub-scheduling_globecom2019} and the main result is depicted in Figure \ref{fig:parallel_sequential}. The figure shows the long term efficiency (expected sum data rate normalized to the capacity of the channel) of the proposed parallel approach, i.e., the transmission of side information and encrypted data are done simultaneously on $\mathcal{\bar{D}}$ and $\mathcal{D}$, respectively, versus a standard sequential transmission approach. It can be seen that, for most values of $\beta$, the parallel approach outperforms the sequential one. Another observations is that as $\beta$ increases, the efficiency decreases. This is expected result as higher $\beta$ will required more frequent key generation, hence, less data transmission. Finally, an important result that can be observed on the graph is that the authors proposed a simple heuristic approach for the parallel scheme that gives an equivalent efficiency to the optimal solution solved using dynamic programming approach (i.e., as a Knapsack problem). Further interesting aspects that can be included in this analysis are factors such as handover or other aspects that may cause frequent key generation.

This problem has been further investigated in \cite{Mitev_Eurasip2020}, where a general quality of service (QoS) delay constraint was introduced. The work is based on leveraging the theory of the effective capacity and identifies the maximum supported transmission rate when considering a delay constraints, i.e., instead of maximizing the data rate $C_D$ the problem focuses on maximizing the effective data rate $E_C (\alpha)$, given as
\begin{equation}
    E_C(\alpha)=-\frac{1}{\alpha} \log_2 \left(\mathbb{E}\left[e^{-\alpha C_D} \right] \right), \label{eq:eff_capacity}
\end{equation}
where $\alpha=\frac{\theta T_f B}{\ln(2)}$ with $\theta$ being a MAC sub-layer parameter that captures the packet arrival rate and introduces a delay requirement into the problem, $T_f$ is the frame duration and $B$ denotes the bandwidth. Considering that, \cite{Mitev_Eurasip2020} identified the optimal power allocation policy that maximizes $E_C(\alpha)$ as 
\begin{equation}
    p_i^*=\frac{1}{g_0^{\frac{N}{\alpha+N}}\hat{g}_i^{\frac{\alpha}{\alpha+N}}}- \frac{1}{\hat{g}_i}, \label{eq:p_star}
\end{equation}
where $g_0$ is a cut-off value that can be found from the power constraint and $\hat{g}_i$ $i=1,\dots, N$ denote the imperfectly estimated channel gains. If the system can tolerate looser delay requirements, i.e., $\theta \rightarrow 0$ the result above converges to the well-known water-filling algorithm and if  stringent delay constraints are implied, i.e., $\theta \rightarrow \infty$ the optimal power allocation converges to total channel inversion. Similarly to the previous case, it has been demonstrated that the parallel approach outperforms the sequential approach, in terms of efficiency, regardless of the values of $\theta$ and $\beta$~\cite{Mitev_Eurasip2020}.

\subsection{Secret key generation under active attacks}

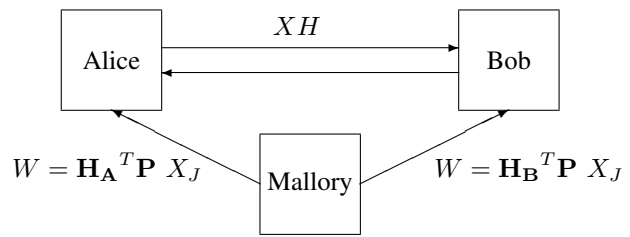
\begin{figure}[!t]
\setlength{\unitlength}{0.13in} 
\centering 
\begin{picture}(23.5,11) 
\put(2,6){\framebox(4,4){Alice}}
\put(10,1){\framebox(4,4){Mallory}}
\put(18,6){\framebox(4,4){Bob}}
\put(6,8.5){\vector(1,0){12}}
\put(10.5,9) {{$XH$}}
 \put(18,7.5){\vector(-1,0){12}}
 \put(11,7.5) {}
\put(10,3){\vector(-2,1){6}}
\put(0,3.3) {{$W=\mathbf{H_A}^T \mathbf{P} \ X_J$}}   
\put(14,3){\vector(2,1){6}}
\put(17,3.3) {$W=\mathbf{H_B}^T\mathbf{P} \ X_J$}
\end{picture}
\caption{Alice and Bob have single transmit and receive antennas and exchange pilot signals $X$ over a Rayleigh fading channel $H$. A MiM, Mallory, with multiple transmit antennas injects a pre-coded signal $\mathbf{P} X_J$, such that the received signals at Alice and Bob are equal $W=\mathbf{H_A}^T \mathbf{P}= \mathbf{H_B}^T\mathbf{P}$.}
\label{fig:injection}
\end{figure}

The previous section discussed how SKG can be used to build authenticated encryption protocols. However, the above scheme could only be secure under the assumption that the advantage distillation phase is robust against active attacks. Therefore, this section focuses on active attacks during SKG, in particular the injection attack is investigated. The idea of this attack is illustrated in Figure \ref{fig:injection}.

Differently from previous sections, instead of an eavesdropper, an active man-in-the-middle (MiM) attacker is considered, referred to as Mallory. The system model assumes two legitimate users, Alice and Bob, each having a single antenna and Mallory, who has two antennas. The goal of the attacker is to inject an equivalent signal $W$ at  both,  Alice and Bob, such that their channel observations $Z_A$ and $Z_B$, respectively, will also include the injected signal: 
\begin{align}
Z_A=XH+W+N_A \label{eq:Za}\\
Z_B=XH+W+N_B, \label{eq:Zb}
\end{align}
where the channel realization between Alice-Bob is denoted by  $H\sim \mathcal{CN}(0, \sigma^2)$, the exchanged signal over this channel is given as $X$, $\mathbb{E}[|X|^2]\leq P$, the noise observations at Alice and Bob are given as $ N_A, N_B \sim \mathcal{ CN}\left( 0, 1 \right)$ and the injected signals over the link Eve-Alice (given as $\mathbf{H}_A$) and Eve-Bob (given as $\mathbf{H}_B$) are given as $W=\mathbf{H_A}^T\mathbf{P} X_J=\mathbf{H_B}^T\mathbf{P}X_J$. The received signals are equal, thanks to the precoding matrix $\mathbf{P}$. A simple mathematical operation can reveal that, as long as Mallory has one extra antenna, as compared to Alice and Bob, the design of the pre-coding matrix is straight forward, i.e., 
\begin{align}
    \mathbf{H_A}^T\mathbf{P} X_J&=\mathbf{H_B}^T\mathbf{P}X_J
    \Rightarrow\nonumber\\ P_1&=\frac{H_{B2}-H_{A2}}{H_{A1}-H_{B1}}P_2. \label{eq:precoding_matrix}
\end{align}

Overall, this is a simple attack to mount its consequences are crucial. As it can be seen in Equations \eqref{eq:Za} and \eqref{eq:Zb}, by injecting the signals, Mallory adds additional term to the shared randomness between Alice and Bob, turning it into $XH+W$. Hence, this allows Mallory to obtain partial information with respect to the generated key.

Fortunately, a simplistic countermeasure has been proposed in \cite{Mitev_globecom_2019_MiM}. The idea is instead of using deterministic pilot signals $X$, as described above, Alice and Bob can transmit independent and randomized probe signals $X$ and $Y$, respectively. This turns their observations into
\begin{align}
Z_A=YH+W+N_A, \label{eq:Za_random}\\
Z_B=XH+W+N_B, \label{eq:Zb_random}
\end{align}
which allows them to simply post-multiply by their own transmission resulting into the following: 
\begin{eqnarray}
\tilde{Z}_A&=&X Z_A=XYH+XW+XN_A,\\
\tilde{Z}_B&=&YZ_B=XYH+YW+YN_B,
\end{eqnarray}
where, as it can be seen, $W$ is not anymore part of the shared randomness. Therefore, as long as $X$ and $Y$ are uncorrelated this simple approach can successfully reduce an injection attack to a less harmful uncorrelated jamming attack. In detail, the jamming attack has impact on the achievable key rate but does not reveal anything about the key to Mallory.

Now, when Mallory's attack is reduced to jamming, a smart thing she can do, is to act as a reactive jammer. A reactive jammer would first sense the spectrum and jam only subcarriers where she detects a transmission. Considering a multicarrier system, Mallory can choose a sensing threshold  and jam only subcarriers where she detects signals with power greater than the chosen threshold. A thorough analysis considering this scenario has be performed in \cite{Mitev_globecom_2019_MiM}, where this problem has been investigated using game theory. In fact, the scenario can be formulated as a non-cooperative zero-sum game with two players, i.e., player $L$, (legitimate users act as a single player), and player $J$, (the jammer). Based on the fact that player $J$ jams only after observing the action from player $L$, this is formed as a hierarchical game with $L$ being the leader of the game and $J$ being the follower. Note that in hierarchical games, the optimal action is the Stackelberg equilibrium (SE). What was shown in this study is that the SE is based on two things: i) the sensitivity of the receiver at player $J$, and more specifically how well the sensing threshold is chosen, and ii) the available power at the legitimate users. The SE is defined as:
\begin{itemize}
    \item If the jammer has badly chosen threshold, depending on the available power at the legitimate users they would optimally:
    \begin{enumerate}
        \item equally distribute their power below the sensing threshold and do not comprise their communication.
        \item transmit with full power on all subcarriers, hence being sensed and jammed.
    \end{enumerate} 
    \item If the jammer has chosen a low threshold that allows to detect all ongoing transmissions, Alice and Bob have no choice but to transmit at full power.
\end{itemize}
Overall, SKG is a promising PLS technology and could help solving the key distribution issue for emerging 6G applications, e.g., addressing scalability for massive IoT~\cite{BITS_Mitev2022}.
% While SKG is a mature technique, a major challenge comes from the fact that SKG rates depend on the variability of channel statistics~\cite{MITEV_vtc2022, MITEV_globecom2022}. Therefore, important research directions on the topic of SKG include: how to access and process channel information, can we trust the information obtained from different chipsets and how does the SKG rate depend on channel parameters.  
%%%%%%%%%%%%%%%%%%%%%%%%%%%%%%%%%%%%%%%%%%%%%%%%%%%%%%%%%%%%%%%%%%%%%%%%%%%%%%%%%
%%%%%%%%%%%%%%%%%%%%%%%%%%%%%%%%%%%%%%%%%%%%%%%%%%%%%%%%%%%%%%%%%%%%%%%%%%%%%%%%%
\section{Authentication Using PLS} \label{sec:Authentication}
%%%%%%%%%%%%%%%%%%%%%%%%%%%%%%%%%%%%%%%%%%%%%%%%%%%%%%%%%%%%%%%%%%%%%%%%%%%%%%%%%
%%%%%%%%%%%%%%%%%%%%%%%%%%%%%%%%%%%%%%%%%%%%%%%%%%%%%%%%%%%%%%%%%%%%%%%%%%%%%%%%%

One of the main motivations to look at PLS authentication schemes is the increasing complexity of standard crypto schemes. In fact, it has been shown in multiple studies that there exists a trade-off between delay and key sizes used in the  cryptographic schemes.
% For example, in \cite{Teniou18}, it has been shown that on a smart vehicle that hosts a $400$ MHz processor, it takes approximately $20$ ms to verify a digital signature. This brings the question whether current standards can comply with systems as ultra reliable low latency communications, where latency requirements are as low as $1$ ms. 

A particular example that focuses on addressing such issues is the zero-round-trip-time (0-RTT) protocol introduced in the TLS version $1.3$ for session resumption. The idea is based on using  resumption keys to quickly resume a session, in a 0-RTT, as opposed to re-authenticating users every subsequent session. Unfortunately, it has been shown that this scheme is vulnerable a set of attacks (e.g., replay attack), however, the community answer was ``But too big a win not to do''~\cite{0rtt_quote}. 

This section gives a hint on what PLS can do in terms of authentication for 6G systems. In particular, it first gives a brief background on physical unclonable functions (PUFs), then discusses how localization can be used as an authentication factor, and finally, it introduces a secure 0-RTT authentication protocol that leverages multiple PLS technologies.

%%%%%%%%%%%%%%%%%%%%%%%%%%%%%%%%%%%%%%%%%%%%%%%%%%%%%%%%%%%%
%%%%%%%%%%%%%%%%%%%%%%%%%%%%%%%%%%%%%%%%%%%%%%%%%%%%%%%%%%%%
\subsection{Physical unclonable functions} \label{sec:PUF}
%%%%%%%%%%%%%%%%%%%%%%%%%%%%%%%%%%%%%%%%%%%%%%%%%%%%%%%%%%%%
%%%%%%%%%%%%%%%%%%%%%%%%%%%%%%%%%%%%%%%%%%%%%%%%%%%%%%%%%%%%

PUFs can be referred to as device fingerprints. The idea is that, the manufacturing of a circuit is a process with unique characteristics (e.g., due to change in the temperature,  vibrations), which makes each device unique on its own. While devices operate in a similar manner, they always have small variations in terms of delays, power-on-state, jitter, etc. This gives an opportunity to leverage these uniqueness, and use it for authentication. 

Given that, a standard PUF based authentication protocol follows two phases. An enrolment phase which takes place offline, and an authentication phase which is performed online. During the enrolment phase, a set of challenges are run on a device's PUF. A set of challenge could refer to measuring propagation delays over different propagation paths. Due to the presence of noise, these measurements are passed through a suitable encoder to generate helper data. Following that, a verifier (e.g., a server) creates a database where challenge-response pairs (CRPs) are stored along with the corresponding helper data. Next, during the online authentication phase, the verifier sends a random challenge to the device, and the device replies with a new PUF measurement. The authentication is successful if the verifier can regenerate the response saved during enrolment by using the new response and the helper data in its database. Note that, to avoid replay type of attacks a CRP should not be re-used. A major advantage of the scheme above is that the device does not need to store any key information and relies only on PUF measurements. Hence, if the device is compromised (e.g., ``captured by an enemy"), no useful information can be extracted.

% Now how can we combine different elements that we discussed so far in order to speed things up for example by introducing 0-round trip time protocols. Actually today the 0-RTT resumptions secret key rs are usually generated by using Diffie Hellman procedures, Diffie Hellman [INAUDIBLE]. 

% So our proposal here is to replace the Diffie Hellman with an SKG step. So the quantized r outputs instead of just having a rA, rB, for the reconciliation. We take into account the resumption secret that will be used for the resumption of the session. So instead of reconciling to rA and rB at the SKG step, we will reconcile today rA x over s and/or rB x over s, and this very simple combination of PUFs for authentication, and SKG for the generation of resumption keys actually allows in a very straightforward manner to build genuine 0-round trip time authentication protocols at the physical layer without any intervention of any cryptographic scheme here. 

\subsection{Location-based authentication }

% \textcolor{red}{Shouldn't the title here be location based as we do not say much about RF, YES}
Localization precision is continuously increasing and the goal of 6G technologies is to achieve centimeter level accuracy. Popular approaches for fingerprinting rely on measuring received signal strength (RSS), carrier frequency offsets, I-Q imbalances, CSI measurements and more. This section presents a lightweight example for location based authentication, through a low-complexity proximity estimation. 

Consider a mobile low-end device with a single antenna and low computational power. Assume that the device has a map of a premise and knows the location of the access points within this premise. A simple strategy to perform reverse authentication (i.e., the device authenticates an access point) is to move in an unpredictable manner and measure the RSS from multiple positions. As the RSS is strongly related to the distance between devices, this simple approach allows to confirm the location of the access point. Typically, localization would require either the deployment of multiple nodes that measure the RSS simultaneously or advanced hardware/computational capabilities when considering a single device. The approach above does not have such requirements and can still be used as an authentication factor. In fact, the proximity detection described above can provide resilience to impersonation type of attacks, e.g., in the presence of a malicious access point. 

% So we perform the proximity estimation in different locations. The received signal strength is non-linear, so there is no way and these locations are unpredictable, so there's no way for a malicious access point to succeed in this attack. And then we can perform proximity estimation and confirm whether, for example the access point is at the expected distance or not. So here we have a simple example for implementing the scheme by using Kalman filters. And as we increase in the distance we see how the output of the Kalman filter actually provides a good estimation of the proximity. 

Now, we summarize some open research issues in the direction of using fingerprint based authentication. A concern that naturally arises is about the resilience of such schemes to jamming and man-in-the-middle type of attacks. In particular, how to cope with interference transmissions, or pilot contamination type of attacks, both of which can alter the precision of the localization information. Another issue concerns the trustworthiness of the localization information, i.e., depending whether we operate at short or long distance, the variability of measurements can change, hence, bringing uncertainty into the system. Finally, another aspect concerns the type of application where such approach could be useful. A good example comes from the idea presented above, e.g., reverse authentication. Reverse authentication can help in mitigating attacks that fall into the general category of false base station attacks (which are open issues in 5G). However, we note that before deploying location-based authentication technologies all concerns must be addressed.

\subsection{Multi-factor PLS authentication} 

A recent publication \cite{Mitev2022_access}, has shown how three PLS credentials (PUFs, SKG and location fingerprints) can be combined into a multi-factor PLS based authentication protocol.
% In detail, i) PUFs are used as a main authentication factor that provide mutual authentication between a mobile node (Alice) and a static server (Bob), ii) proximity estimation is used by Alice as a second authentication factor, to re-assure her for the legitimacy of Bob, and, iii) SKG is used to randomize resumption keys offering a forward security and resilience to replay attacks on 0-RTT authentication. 
The proposed scheme uses PUFs as a mutual authentication factor between a mobile node (Alice) and a static server (Bob). The protocol is realized following a typical PUF approach, i.e., following two steps, enrolment and authentication. The use of PUFs provides several security guaranties, including protection against physical and cloning attacks. Next, Alice uses proximity estimation as a second authentication factor. This simple technique re-assures her for the legitimacy of Bob and provides resistance to impersonation attacks (e.g., false base station attacks). To provide anonymity for Alice, the scheme introduces one-time alias IDs. After a successful authentication, both parties exchange resumption secrets, following a standard TLS 1.3 procedure. The resumption secrets are used for a fast 0-RTT re-authentication between Alice and Bob, i.e., session resumption (as opposed to performing a full authentication procedure). While the standard approach for session resumption is not forward secure and is vulnerable to replay attacks, the scheme in~\cite{Mitev2022_access} uses SKG keys to randomize the resumption secrets. It is shown that adding SKG ensures both perfect forward security and resistance against replay attacks. 

In general, using the physical layer for authentication is a well investigated topic. Schemes like the one above, show that there are already multiple PHY schemes which can contribute for the system security. Some of the research problems in the area include  design of high-entropy PUFs and accurate and privacy-preserving location-based authentication.
\section{Conclusions and Future Directions} \label{sec:Prospects}
%%%%%%%%%%%%%%%%%%%%%%%%%%%%%%%%%%%%%%%%%%%%%%%%%%%
%%%%%%%%%%%%%%%%%%%%%%%%%%%%%%%%%%%%%%%%%%%%%%%%%%%

This paper highlights the role that PLS could play in 6G, in view of the evolution in terms of security, with the concepts of trust, context awareness, and quality of security. 

6G is expected to introduce new features to communication standards including sensing, subTHz communication, massive MIMO, extreme beamforming, learning and actuating, ultra reliable low latency computing and more.
% Combining these features together could bring the remote control for mission critical IoT and enable ultra reliable low latency transmissions and decisions. 
While it is still not clear how the transition from 5G to 6G will look like, there is growing interest on the use of semantics, semantic communications, semantic compression, and context awareness in 6G.
% As an example, in \cite{popovski_semanticFiltering} the use of semantic filtering in the post 5G connectivity era was discussed. The idea is to introduce a new plane, the so-called semantic plane, that acts as a translator between different layers. This is where semantic compression is performed and it can help in achieving all low latency communications and computations aspects that were discussed above.

Another perspective was introduced
% based on network slicing. Network slicing brought the need to introduce the term 
with quality of security (QoSec),
% in service level agreements (SLAs). 
i.e., different slices of the network have different security and privacy requirements. This brings the need of adaptive security levels.
% As an example, in a 0-trust scenario in order to achieve quantum level of security the security level is said to be level $5$. However, other levels are currently not defined.
A series of questions arise based on the above: How to define other security levels? How to perform adaptive identity management? How to make an intelligent risk assessment?

PLS emerges as a contestant for the next generation of security systems in 6G. One key advantage of PLS is that it is inherently adaptive. This is due to the fact that in physical technologies, the secrecy outage probability can be directly tuned through adjusting the transmission rate.
In particular, wireless channels can be treated as a source of two things, a source of uniqueness, and a source of entropy. For example, in a slow flat fading scenario (e.g. LoS) then the channel could be treated as a good source of uniqueness. As discussed in Section \ref{sec:Authentication}, uniqueness can be easily used for authentication purposes. On the other hand, if the channel changes very fast, due to small scale fading, it could be treated as a good source of entropy. The variability of the channel can then be directly used to either distill keys, or perform keyless transmission. An important observation is that if one is not available, e.g., uniqueness, then the other will be, e.g., entropy. 

Following the above, an open research question is, how to characterize the channel properties and particularly, which part of the channel should be considered as predictable and which as unpredictable. It is not an easy question to answer as it would require the characterization of the channel correlations in time, frequency and space domains; but it is an important one as it would allow the alignment of PLS metrics to semantic security metrics.

% So what else can the physical layer do for 6G security? So we can leverage context awareness in 6G security. Pretty much what I talked about. Learning the channel, learning the infrastructure, understanding the physical parameters, so is this massive MIMO, is this a UAV type of network, what type of a network is it, can I learn the statistics of the wireless medium and of the topology? 

% And then hopefully I can use physical security to adapt the security level. If I do it correctly, I have a very cheap means, a very cheap way of providing both quantum guarantees. And all of this in an adaptive QoSec framework. And there are new possibilities with new PHY networking aspects, for example in Terahertz communications we can think of extreme beamforming scenarios where basically we'll expect to have, I would say RF, the equivalent of RF wire. So very thin beams. 

% And this can provide it's an open question, but it can possibly provide a case for the wire tap channel to be for the applications of the wire tap channel. Or we can leverage for example, visible light communications technologies and these are very hard to eavesdrop naturally. We can think of new learning opportunities through graph signal processing and network tomography. Or we can look at satellite communications where actually the application of quantum key distribution is expected to take place in essence. 

Finally, we believe it is now time to start defining the security levels based on the usage of multiple elements. Here, we list several elements:

\begin{enumerate}
    \item Criticality of information - how important the information is from user or the network perspective;
    \item Value of information for the attacker - this captures, who is the attacker and how much effort is expected to put into compromising the system;
    \item System resilience - this includes the stability and repair time after an attack;
    \item Threat level - the usage of context to recognize ``abnormal'' events (could include location, behavior and communication information);
    \item QoS constraint - systems are expected to comply with particular QoS index.
    \end{enumerate}

Today, the deployment of PLS in systems is still lacking traction. However, there is a growing interest by industry and academia. This paper shows the potential of PLS for upcoming wireless system designs. It gives concrete examples of use cases for PLS, reaching far beyond addressing encryption. By doing so, greatly improving the security of 6G networks. For PLS it is instrumental to characterize and exploit the wireless channel from a security point of view. A key advantage is seen for developing light-weight security solutions for low-latency and massive IoT use cases.

\section*{Acknowledgement}

This work is financed by the Saxon State government out of the State budget approved by the Saxon State Parliament.
\bibliographystyle{IEEEtran}

\bibliography{reference}

\end{document}